\documentclass[final,3p,times]{elsarticle}
\usepackage{amssymb,amsmath}
\journal{Physica A: Statistical Mechanics and its Applications}

\newcommand{\mean}{\mathsf{E}}

\begin{document}

\begin{frontmatter}



\title{The influence of the finite velocity on spatial distribution of particles in the frame of Levy walk model\tnoteref{t1}.}
\tnotetext[t1]{This work was supported by the Ministry of Education and Science of the Russian Federation (No. 6.1617.2014/K).}


\author[svv]{Viacheslav V. Saenko}
\ead{saenkovv@gmail.com}
\address[svv]{Technological Research Institute S.P. Kapitsa, Ulyanovsk State University,Leo Tolstoy str. 42, Ulyanovsk, Russia, 432000}

\begin{abstract}
Levy walk at the finite velocity is considered.  To analyze the spatial and temporal characteristics of this process, the method of moments has been used. The asymptotic distributions of the moments (at $t\to\infty$) have been obtained for  $N$ dimensional case where the free path of particles demonstrates the power-law distribution $p_{\xi}(x)=\alpha x_0^\alpha x^{-\alpha-1}$, $x\to\infty$, $0<\alpha<2$. The three regimes of distribution have been distinguished:  ballistic, diffusion and asymptotic. Introduction of the finite velocity requires considering of two problems: propagation with distribution at the finite mathematical expectation of the free path ($1<\alpha<2$) and propagation with distribution at the infinite mathematical expectation of the free path of the particle ($0<\alpha<1$). In the case $1<\alpha<2$, the asymptotic distribution is described by the Levy stable law and the effect of the finite velocity is reduced to a decrease of diffusivity. At $0<\alpha<1$, the situation is quite different. Here, the asymptotic distribution exhibits a $U$- or $W$-shape and is described as the ballistic regime of distribution. The obtained moments allow to reconstruct the distribution densities of particles in one-dimensional and three-dimensional cases.
\end{abstract}

\end{frontmatter}

%\linenumbers

Here, we consider an effect of the finite velocity on spatial distribution of the particles at anomalous diffusion. It is well known that the anomalous diffusion is defined by the power-law dependence of the diffusion packet width on time $\Delta (t)\propto D_\alpha t^\gamma$,where $D_\alpha$ is the diffusion coefficient \cite{Metzler1994, Fogedby1994a,Fogedby1994, Shlesinger1987,Klafter1987}. Different regimes of this process are recognized depending on the exponent value $\gamma$: normal diffusion ($\gamma=1/2$), subdiffusion ($\gamma<1/2$), and superdiffusion ($\gamma>1/2$). At $\gamma=1$ and $\gamma>1$, the quasi-ballistic and superballistic regimes, respectively, are established. For more details on each regimes see Refs.~\cite{Chukbar1995,Zolotarev1999,Uchaikin1999a,Zaburdaev2002}.

Anomalous diffusion is essential for study due to a wide range of its application. Anomalous diffusion is known for relaxation processes in dielectrics \cite{Nigmatullin1984}, turbulent fluxes of the particles at the edge of the plasma cord \cite{Carreras1999b, Trasarti-Battistoni2002, Carreras1998,Saenko2010, Saenko2009, Skvortsova2005}, central region of plasma cord \cite{Hauff2007} in closed magnetic traps,  study of mRNA molecule diffusion in cells \cite{Burnecki2012} , geological and geophysical processes, and  biological systems (see \cite{Metzler2004} and Refs. there). The anomalous diffusion models are employed to describe propagation of cosmic rays \cite{Ragot1997,Zimbardo2012, Duffy1995, Perri2007,Perri2008a,Perri2009} and their acceleration \cite{Bian2008,Perri2012,Lazarian2013,Zimbardo2013}, to study wandering of interstellar magnetic field lines \cite{Ragot2011,Ragot2009,Shalchi2011,Ragot2013}, to describe heat transfer in the systems which do not obey the Fourier conductivity law \cite{Dhar2013}, to develop dynamic models describing sequences of nucleotides in DNA molecule \cite{Allegrini1995}.

The Continuous Time Random Walk (CTRW) model is used to describe anomalous diffusion \cite{Montroll1965,Scher1973,Scher1973a,Scher1975,Pfister1978,Klafter1980,Klafter1987}. In order to take into account the finite velocity, the variations of CTRW model have been introduced. Conventionally these modifications are classified into two groups: 1) coupled Levy walk and 2) velocity Levy walk. The first group models assume a walk to be sequence of instantaneous jumps followed by rest sate, while, in the second group models a particle moves continuously between two successive collisions. The CTRW model differs from the coupled Levy walk in free path dependence on resting time. The dependence of these distributions is described by the transition density $\psi(r,t)$. Ref. \cite{Shlesinger1982} reports conditions at which $\psi(r,t)$ does not expand into two independent factors. It has been shown that the main condition is the power-law distributions of walk and the resting time. Noteworthy, in the case of CTRW process, this density is multiplication of two independent factors $\psi(r,t)=p_\xi(r)q_\tau(t)$  where $p_\xi(r)$ and $q_\tau(t)$ are free paths and rest time distributions respectively. The following transition density has been used in Ref. \cite{Klafter1987} $\psi(x,t)=Cx^{-\mu}\delta(x-t^\nu)$. The authors state that this function allows a particle to make an arbitrary long walk, but longer walks require longer time. Further, coupled Levy walks with the similar transition density $\psi(r,t)$ have been studied in Refs. \cite{Blumen1989,Zumofen1989,Wang1992}. Ref. \cite{Zaburdaev2006} reports modified coupled Levy walks for strong dependence of the resting time on the preceding jump length. Here, the effects of different dependences on the asymptotics of the process are studied. Study of the standard deviation of diffusing particles and self-similar properties of distributions of various modifications of models based on transition core is reported in Refs. \cite{Schmiedeberg2009,Liu2013}. Limit distributions and fractional differential equations describing this process are studied in Refs. \cite{Becker-Kern2004,Froemberg2015} and \cite{Jurlewicz2012}, respectively.

The second group models (velocity Levy walk) are based on the assumption that a particle continuously changes its coordinates in motion. One of the main reasons for introducing the finite velocity is consideration of physical problems in which the hopping-trap mechanism does not allow to describe the physical nature of process. This modification of the anomalous diffusion model referred to as "Levy walk" is used in Ref. \cite{Shlesinger1987} for description of turbulence. Ref. \cite{Shlesinger1985} reports that the model of rotation phase dynamics in Josepson junctions \cite{Geisel1985} leads to anomalous diffusion. Further, this model has been studied in works \cite{Zumofen1993, Klafter1993, KlafterJ1994}. Levy walks in bounded and semibounded space are studied in \cite{Drysdale1998}. Ref. \cite{Uchaikin1998} reports kinetic equations of anomalous diffusion at the finite speed, standard deviation of particles and exact solution for one-dimensional case. Paper \cite{Andersen2000} is devoted to statistical moments for Levy walks at the finite velocity without traps in one-dimensional case. Multidimensional walks with traps of an arbitrary type at the finite velocity are investigated in \cite{Uchaikin_TMF1998b_eng, JarovikovaPhD2001_en, Uchaikin2003a}. These works confirm the results of Ref. \cite{Zolotarev1999} where the authors consider influence of the finite velocity on spatial distribution of particles in the frame of Levy walks with the traps of exponential type. Here, it has been reported that at $1<\alpha\leqslant2$ the accounting of the finite velocity is reduced to renormalization of the diffusion coefficient in the fractional diffusion equation and solution of the equation has been expressed through the Levy stable law. At $0<\alpha\leqslant1$, the fractional diffusion equation cannot describe the process, since the asymptotic distributions are $U$- and $W$-shaped. The same conclusion has been drawn in Ref.~\cite{Ferrari2001}, where the generalization of telegraph model has been applied to the case of Levy walks. Authors of Refs.~\cite{Zaburdaev2008,Froemberg2015} study Levy walks at the finite velocity without traps and state that at the infinite mathematical expectation of free paths the asymptotic distributions of particles demonstrate $U$-shape and $W$-shape.

Further, Levy walks at the finite velocity have been studied in Ref.~\cite{Zaburdaev2002}. Authors show that at the infinite mathematical expectation of free paths taking into consideration of the finite velocity leads not only to decrease of the diffusion constant in fractional diffusion equation as reported in Ref.~\cite{Zolotarev1999} but also to replacement of the fractional Laplace operator with material derivative of fractional order in fractional diffusion equation. Further development of the idea of material derivative of fractional order introduced into consideration of Levy walks at the finite speed has been provided in Refs.~\cite{Zaburdaev2006,Sokolov2003,Chukbar2003,Uchaikin:JETPL:2010,Uchaikin2011}. This operator is applied for description of cosmic ray propagation in the Galaxy \cite{Uchaikin2011}, for study of the propagation of resonance radiation \cite{Uchaikin2011a}, for description of cosmic ray propagation along the magnetic field lines in the Solar system \cite{Uchaikin2014}. A wide range of applications of the material derivative stimulated progress in solving of the fractional diffusion equation using this operator. Solution of the fractional diffusion equation with the material derivative of fractional order expressed through the Lamperti distribution (see also \cite{Uchaikin2009}) has been obtained in Ref.~\cite{Uchaikin2011} for one-dimensional walks without traps. Similar limiting distribution has been derived in Ref.~\cite{Froemberg2013a}. Here, random walks are considered at the constant velocity $v_0$ and power-law distribution of free paths. Authors reveal that at the infinite mathematical expectation of free paths the limiting distribution is described by the Lamperti distribution, while at the finite mathematical expectation – by the Levy stable law.

Thus accounting of the finite speed at the infinite mathematical expectation of free paths essentially changes not only the fractional diffusion equation but also asymptotic distribution of the particles. However, Refs.~\cite{Zaburdaev2006, Schmiedeberg2009, Froemberg2015, Magdziarz2012, Meerschaert2002b} report similar results obtained under certain conditions for coupled Levy walks and Levy walks at the finite velocity. In particular, if the resting time in the coupled Levy walk model demonstrates a linear dependence on the length of preceding jump, this model can be described by the same equation \cite{Zaburdaev2006,Meerschaert2002b} and exhibits the same root mean square deviation of particles \cite{Schmiedeberg2009} as in the velocity Levy walk model. However, the asymptotic distributions are different \cite{Froemberg2015,Magdziarz2012}. A more detailed description of the approaches describing Levy walks and systems where they appear is given in review \cite{Zaburdaev2014}.

In the present paper, we study influence of the finite velocity of free motion in the frame of velocity Levy walks (random walks at the finite velocity without traps) on asymptotic distribution of particles. Importantly, similar problem has been developed in Refs.~\cite{JarovikovaPhD2001_en,Uchaikin2003a}. In these papers the velocity Levy walks with traps have been considered and asymptotic distribution of particles has been reconstructed using statistical moments. However here, the expressions for the main asymptotic terms of moments only have been derived.  Besides, in Ref.~\cite{JarovikovaPhD2001_en} the authors have concluded that the $W$-shape of asymptotic distribution is due to predominance of particle capturing by traps over free motion of particles. However, this conclusion contradicts the statement given in this paper that at ($0<\alpha\leqslant1$) the $W$-shape distribution is obtained even without traps.

The work is organized as follows. In Sec.~\ref{sec:momeq} we describe the model of random walk and derive the basic equation for statistic moments. In Sec.~\ref{sec:momAsymp} the asymptotic distributions are obtained for moments assuming the power-law distribution of free paths and followed by the analysis of the moments given in Sec.~\ref{seq:momAnalysis}. In Sec.~\ref{sec:PDFReconstr} we reconstruct the asymptotic distribution using orthogonal polynomials. In Sec.~\ref{sec:Koeff} the expression for anomalous diffusion constant is derived. The obtained results are discussed in Sec.~\ref{sec:disscusion}.

\section{Moment equations}\label{sec:momeq}

Our task is to determinate a spatial distribution of the particles and study the effect of the finite velocity on its asymptotic (at $t\to\infty$) behavior. We use the method of moments developed in fixed velocity charge transfer theory for cases of normal \cite{Yarovikova2000eng_1, Yarovikova2000eng_2} and anomalous  \cite{JarovikovaPhD2001_en,Uchaikin2003a} diffusion. The application of the method of moments is permitted due to a finite velocity of propagation. Indeed, a finite velocity means that for the time $t$ the particle cannot leave the area $\mathbf{R}=\mathbf{v}t$, so all spatial moments of the distribution are possible.

To determine the spatial moments one can apply the method used in Refs \cite{JarovikovaPhD2001_en,Uchaikin2003a}, where more general results for walks with a finite velocity and traps are obtained. However, there are some mistakes in these works. As an example, substituting the expression for $g_{2j}(\lambda)$ (see, section 3 after (23) in Ref.~\cite{Uchaikin2003a} or section 2.1 after (2.1.5) in Ref.~\cite{JarovikovaPhD2001_en}) into  $\tilde\mu^{\langle2n\rangle}(\lambda)$ (see, (24) in Ref.~\cite{Uchaikin2003a} or (2.1.6) in Ref.~\cite{JarovikovaPhD2001_en}), we come to expression for Laplace image of even moments that is wrong, so the expression for $g_{2j}(\lambda)$ should be verified. Verification of this expression forces us to fully repeat all calculations in the model considered in this work and then to eliminate traps from consideration. It is beyond the scope of this work. Therefore to find moments of the distribution we use the method proposed in Ref.~\cite{Uchaikin_TMF1998b_eng} that has been already successfully applied for description of spatial moments in the case of normal diffusion \cite{Yarovikova2000eng_1} and for reconstruction of the spatial distribution of particles through orthogonal polynomials \cite{Yarovikova2000eng_2}.

Let us consider the symmetric walk in $N$-dimensional space. Let $\mathbf{R}_N(t)$ be $N$-dimensional random vector characterizing displacement of the particle for time $t$. The particle velocity is considered to be constant and independent of the direction of motion. We assume that diffusion is isotropic and the particle source is a point source. With these assumptions the following stochastic relationship can be written for the random displacement vector:
\begin{equation}\label{eq:R}
\mathbf{R}_N(t)=\left\{\begin{array}{ll}
vt\mathbf{\Omega},\ \mbox{with the probability}\ P_\xi(vt)d\Omega_N/S_N,\\
\xi\mathbf{\Omega}+\mathbf{R}_N(t-\xi/v),\ \mbox{with the probability}\ p_\xi(x)dx d\Omega_N/S_N,
\end{array}\right.
\end{equation}
where $p_\xi(x)dx$ is the probability of particle collision around the point $x$, and
$P_\xi(x)=\int_x^\infty p_\xi(y)dy$, $\mathbf{\Omega}$ is the unit vector, $d\Omega_N$ is the segment of the sphere surface with unit radius centered in $\mathbf{\Omega}$, $S_N$  is the sphere area $S_N=\int d\Omega_N$. Due to space symmetry all odd moments are zero. Taking (\ref{eq:R}) in $2n$- power, where $n=1,2,3,\dots$ we come to
\begin{equation}\label{eq:R2n}
\mathbf{R}_N^{2n}(t)=\left\{\begin{array}{ll}
(vt)^{2n},\ \mbox{with the probability}\ P_\xi(vt)d\Omega_N/S_N,\\
(\xi\mathbf{\Omega}+\mathbf{R}_N(t-\xi/v))^{2n},\ \mbox{with the probability}\ p_\xi(x)dx\, d\Omega_N/S_N,
\end{array}\right.
\end{equation}
Now averaging (\ref{eq:R2n}) over the random variables we expand it to:
\begin{align}\label{eq:m2n}
m_{2n}^N(t)=&(vt)^{2n}P_\xi(vt)+\sum_{n_1+n_2+n_3=n}\hspace{-2mm}2^{n_2}\frac{n!}{n_1!n_2!n_3!}
\int\limits_0^{vt}x^{2n_1+n_2}m_{n_2}^N(t-x/v) m_{2n_3}^N(t-x/v)
\langle\cos^{n_2}\theta\rangle p_\xi(x)dx
\end{align}
where the sum is taken over all equation solutions $n_1+n_2+n_3=n$. Here we introduce $m_{2n}^N(t)\equiv \mean \mathbf{R}_N^{2n}(t)$ the stochastic moment of $2n$-order of the random vector $\mathbf{R}$ in $N$-dimension space, $\mean X$ is the mathematical expectation of the random value $X$.
\begin{equation}\label{eq:intcosn}
\langle\cos^{n}\theta\rangle=\frac{1}{S_N}\int\limits_{\Omega_N}(\cos\theta)^n d\Omega_N
\end{equation}
is the averaged cosine, $\theta$ is the angle between the vectors $\mathbf{R}_N(t-\xi/v)$ and $\mathbf{\Omega}$, (see, \ref{app:cos} for derivation of (\ref{eq:intcosn})). One can see that the expression~(\ref{eq:m2n}) is recurrent, so for determination of the $2n$ order moment we need to know all moments of lower orders.

From the normalization conditions for the distribution density $\int p(x,t)dx=1$ we can get $m_0^{N}=1$. Substituting $n=1$ in (\ref{eq:m2n}) we come to the expression for the second moment:
\begin{equation}\label{eq:m2tmp}
m_2^N(t)=(vt)^2P_\xi(vt)+\int_0^{vt}x^2 p_\xi(x)dx
+\int_0^{vt}m_2^N(t-x/v))p_\xi(x)dx.
\end{equation}
Here, $m_1^N(t)=0$, $(1/S_N)\int_{\Omega_N}\cos\theta d\Omega_N=0$ and $(1/S_N)\int_{\Omega_N} d\Omega_N=1$
are taken into account. Below, we will use the following integral obtained by integration in parts:
\begin{equation}\label{eq:intXiK}
\int_0^{vt}x^k p_\xi(x)dx=-\int_0^{vt}x^k\frac{dP_\xi(x)}{dx}dx
=-(vt)^kP_\xi(vt)+k\int_0^{vt}x^{k-1} P_\xi(x)dx.
\end{equation}
Using (\ref{eq:intXiK}) for integration of the second term in (\ref{eq:m2tmp}) we get the final expression for the second order moment:
\begin{equation}\label{eq:m2}
m_2^N(t)=2\hspace{-1.5mm}\int_0^{vt}\hspace{-2mm}x P_\xi(x)dx+\hspace{-1.5mm}\int_0^{vt}\hspace{-2mm}p_\xi(x) m_2^N(t-x/v)dx.
\end{equation}

For the 4-order moment Eq.(\ref{eq:m2n}) gives
\begin{align}
m_4^N(t)=&(vt)^4P_\xi(vt)+\int\limits_0^{vt}x^4p_\xi(x)dx
+\int\limits_0^{vt}2x^2m_2^N(t-x/v)(1-2\langle\cos^2\theta\rangle)p_\xi(x)dx
+\int\limits_0^{vt}m_4^N(t-x/v)p_\xi(x)dx.\label{eq:m4tmp}
\end{align}
Applying (\ref{eq:meanCos}) for integration and (\ref{eq:intXiK}) for the second term in (\ref{eq:m4tmp}) we come to
\begin{equation}\label{eq:m4}
m_4^N(t)=4\int\limits_0^{vt}x^3P_\xi(x)dx+\int\limits_0^{vt}\left(\tfrac{4+2N}{N}x^2m_2^N(t-x/v)
+m_4^N(t-x/v)\right)p_\xi(x)dx.
\end{equation}
A similar procedure allows to obtain expressions for moments of any order. The equations for $m_6^N(t), m_8^N(t)$ and $m_{10}^N(t)$ are
\begin{align}
m_6^N(t)=&6\int\limits_0^{vt}x^5P_\xi(x)dx+
\int\limits_0^{vt}\left(\tfrac{3N+12}{N}\left(x^4m_2^N(t-x/v)+x^2m_4^N(t-x/v)\right)
+m_6^N(t-x/v)\right)p_\xi(x)dx,\label{eq:m6}\\
m_8^N(t)=&8\int\limits_0^{vt}x^7P_\xi(x)dx +\int\limits_0^{vt}\left[ \tfrac{4N+24}{N}\left(x^6m_2^N(t-x/v)+x^2m_6^N(t-x/v)\right)\right.\nonumber\\
+&\left.\tfrac{6N(N+2)+48(N+3)}{N(N+2)}x^4m_4^N(t-x/v)+m_8^N(t-x/v)\right]p_\xi(x)dx,\label{eq:m8}\\
m_{10}^N(t)=&10\int\limits_0^{vt}x^9P_\xi(x)dx+\int\limits_0^{vt}
\left[\tfrac{5N+40}{N}\left(x^8m_2^N(t-x/v)+x^2m_8^N\hspace{-0.5mm}(t\hspace{-0.5mm}-\hspace{-0.5mm}x/v)\right)\right.\nonumber\\
+&\left.\left(10\hspace{-0.5mm}+\hspace{-0.5mm}\tfrac{120}{N}\hspace{-0.5mm}+\hspace{-0.5mm}\tfrac{240}{N(N+2)}\right)\hspace{-1mm}
\left(x^6m_4^N\hspace{-0.5mm}(t\hspace{-0.5mm}-\hspace{-0.5mm}x/v)+x^4m_6^N(t-x/v)\right)
+m_{10}^N(t-x/v)\right]p_\xi(x)dx\label{eq:m10}
\end{align}

Solutions of equations (\ref{eq:m2}), (\ref{eq:m4}), (\ref{eq:m6}), (\ref{eq:m8}), (\ref{eq:m10}) could be obtained through Laplace transformations. With
$$
\hat m_n(\lambda)=\int\limits_0^{\infty} m_n^N(t)e^{-\lambda t}dt,\quad
\hat p_n(\lambda)=\int\limits_0^{\infty} x^n p_\xi(x)e^{-\lambda x}dx,\quad
\hat P_n(\lambda)=\int\limits_0^{\infty} x^n P_\xi(x)e^{-\lambda x}dx,
$$
the equations for the moments in terms of Laplace images become algebraic that are easy to solve:
\begin{align}
\hat m_2(\lambda)&=\frac{2\hat P_1(\lambda/v)}{\lambda(1-\hat p_0(\lambda/v))},\label{eq:m2laplace}\\
\hat m_4(\lambda)&=\frac{\tfrac{4}{\lambda}\hat P_3(\lambda/v)+\tfrac{4+2N}{N}\hat p_2(\lambda/v)\hat m_2(\lambda)}{1-\hat p_0(\lambda/v)},\\
\hat m_6(\lambda)&=\frac{1}{1-\hat p_0(\lambda/v)}\Big(\tfrac{6}{\lambda}\hat P_5(\lambda/v)
+\tfrac{3N+12}{N}\big( \hat p_4(\lambda/v)\hat m_2(\lambda)+\hat p_2(\lambda/v)\hat m_4(\lambda)\big)\Big),\\
\hat m_8(\lambda)&=\frac{1}{1-\hat p_0(\lambda/v)}\Big(\tfrac{8}{\lambda}\hat P_7(\lambda/v)
+\tfrac{4N+24}{N}\left( \hat p_6(\lambda/v)\hat m_2(\lambda)+\hat p_2(\lambda/v)\hat m_6(\lambda)\right)\nonumber\\
+&\tfrac{6N(N+2)+48(N+3)}{N(N+2)}\hat p_4(\lambda/v)\hat m_4(\lambda)\Big),\\
\hat m_{10}(\lambda)&=\frac{1}{1-\hat p_0(\lambda/v)}\Big(\tfrac{10}{\lambda}\hat P_9(\lambda/v)
+\tfrac{5N+40}{N}\big( \hat p_8(\lambda/v)\hat m_2(\lambda)+\hat p_2(\lambda/v)\hat m_8(\lambda)\big)\nonumber\\
+&\tfrac{10N(N+2)+120(N+4)}{N(N+2)}\big(\hat p_6(\lambda/v)\hat m_4(\lambda)
+\hat p_4(\lambda/v)\hat m_6(\lambda)\big)\Big)\label{eq:m10laplace}
\end{align}

\section{Asymptotic of the moments}\label{sec:momAsymp}

The obtained system (\ref{eq:m2laplace})~-~(\ref{eq:m10laplace}) is rather exact, since no assumptions about the form of the path distribution $p_\xi(x)$ have been done.  No simplifications have been applied deriving set of equations (\ref{eq:m2n}).

In the case of an exponential distribution of paths $p_\xi(x)=\mu\exp(-\mu x)$ the problem is simplifying to the problem of normal diffusion with the finite velocity of free motion. In one-dimensional case, this problem is precisely described by the telegraph equation (see~\cite{Saenko2001_JTP_eng,Uchaikin2000a}). The Laplace images in (\ref{eq:m2laplace}) take the form $\hat p_0(\lambda/v)=\mu v/(\lambda+\mu v), \hat P_1(\lambda/v)=v^2/(\lambda+\mu v)^2$. Substituting these expressions in (\ref{eq:m2laplace}) we obtain $\hat m_2(\lambda)=2v^2/\lambda^2(\lambda+\mu v)$.  Applying an inverse Laplace transformation to this expression, we obtain an expression for the second moment $m_2^D(t)=2(\mu vt-1+\exp(-\mu vt))/\mu^2$ that is exactly the same as obtained in Ref.~\cite{Uchaikin2000a}. Now considering a limit at $t\to\infty$ we get $m_2^D(t)\approx 2Dt$, where $D=v/\mu$, that is the second order moment for normal diffusion with the diffusion coefficient $D$. Note, if the space dimension is greater than one, the telegraph equation becomes approximate (see,~\cite{Uchaikin2000a}), and describes the walk process less accurately than the diffusion equation.

Let us now assume the paths to demonstrate a power-law distribution:
\begin{equation}\label{eq:racePDF}
p_\xi(x)\left\{\begin{array}{ll}
0,&x< x_0\\
\alpha x_0^\alpha x^{-\alpha-1},& x\geqslant x_0,
\end{array}\right.
\end{equation}
where $0<\alpha\leqslant2$. Only moments with orders $k<\alpha$ exist for such kind of a distribution, i.e. $\mean\xi^k<\infty$ at $k<\alpha$  and $\mean\xi^k=\infty$ at $k\geqslant\alpha$.	It means that at  $0<\alpha\leqslant1$ the mathematical expectation is infinity and at $1<\alpha\leqslant2$ the mathematical expectation is finite. These two cases should be considered independently.

\paragraph{Case $0<\alpha<1$} For derivation of spatial moments we need the expressions for $\hat p_0(\lambda), \hat p_n(\lambda), \hat P_n(\lambda)$. For $\hat p_0(\lambda)$ after integration in parts:
$$
\hat p_0(\lambda)=\alpha x_0^\alpha\int_{x_0}^\infty x^{-\alpha-1}e^{-\lambda x}dx=e^{-\lambda x_0}-\lambda x_0^\alpha
\int_{x_0}^{\infty}x^{-\alpha}e^{-\lambda x}dx.
$$
Then, transforming  $y=\lambda x$ we obtain
\begin{equation}\label{eq:p0a1}
\hat p_0(\lambda)=e^{-\lambda x_0}-(\lambda x_0)^\alpha\int_{\lambda x_0}^\infty y^{-\alpha}e^{-y}dy=
e^{-\lambda x_0}-(\lambda x_0)^\alpha\Gamma(1-\alpha,\lambda x_0),
\end{equation}
where $\Gamma(a,x)$ is the incomplete Gamma-function.

According to the Tauber's theorems, the asymptotic behavior at $t\to\infty$ corresponds to the asymptotic behavior of the Laplace image at $\lambda\to0$. Since $\mean\xi=\infty$, expanding the exponent in a series only the terms with $\lambda^k$, where $k<1$ should be kept. As a result $\exp(-\lambda x_0)\approx1$. For the incomplete gamma function $\Gamma(1-\alpha,\lambda x_0)\to \Gamma(1-\alpha)$ at $\lambda\to0$. Substituting these expansions in (\ref{eq:p0a1}) we finally obtain
\begin{equation}\label{eq:p0}
\hat p_0(\lambda)\approx1-(\lambda x_0)^\alpha\Gamma(1-\alpha).
\end{equation}
Similarly, the expressions for other Laplace images could be obtained.
\begin{equation}\label{eq:pn}
\hat p_n(\lambda)=\alpha x_0^\alpha\int_{x_0}^{\infty}x^{n-\alpha-1}e^{-\lambda x}dx=\alpha x_0^{\alpha}\lambda^{\alpha-n}\Gamma(n-\alpha,\lambda x_0)\xrightarrow[\lambda\to0]{} \alpha x_0^{\alpha}\lambda^{\alpha-n}\Gamma(n-\alpha),
\end{equation}
\begin{equation}\label{eq:PnUpper}
\hat P_n(\lambda)=x_0^\alpha\int_0^\infty x^{n-\alpha}e^{-\lambda x}dx=x_0^\alpha\lambda^{\alpha-n-1}\Gamma(n-\alpha+1).
\end{equation}

Substituting (\ref{eq:p0}) and (\ref{eq:PnUpper}) in (\ref{eq:m2laplace}) and simplifying we obtain the expression for the asymptotic function of the second moment $\hat m_2(\lambda)=(1-\alpha)v^2\Gamma(3)\lambda^{-3}$.
Now, applying the inverted Laplace transfer we come to the final expression:
\begin{equation}\label{eq:m2a01}
m_2^N(t)=\left( 1-\alpha \right)(vt)^2,
\end{equation}
On the same way the expressions for other moments are obtained. The expressions for $m_4^N(t), m_6^N(t), m_8^N(t), m_{10}^N(t)$ are listed in \ref{app:MomentsExpr} (\ref{eq:m4a01})~-~(\ref{eq:m10a01}).

Let us introduce the term of diffusion packet width as $\Delta(t)=\sqrt{m_2^N(t)}$. In general case, it increase as $\Delta(t)\propto t^\gamma$. Depending on the parameter $\gamma$ different kinds of diffusion can be obtained: $\gamma>1/2$ superdiffusion, $\gamma<1/2$ - subdiffusion and $\gamma=1/2$ normal diffusion.
In its turn superdiffusion can be classified in three groups: superdiffusion regime at $1/2<\gamma<1$, quasi-ballistic regime at $\gamma=1$ and superballistic regime at $\gamma>1$. The meaning of quasi-ballistic regime is that the diffusion package expands with a speed of free motion of particles; super-ballistic regime corresponds to the expansion of the diffusion packet faster than free motion of particles.

Obtained from (\ref{eq:m2a01}) $\Delta(t)=\sqrt{1-\alpha}vt$ corresponds to quasi-ballistic diffusion, according to the introduced terms. In other words, when the distribution of free paths $p_\xi(x)$ has no mathematical expectation the package expands with a speed of free propagation of particles. It is obvious that the particle propagating with the velocity $v$ for time $t$ passes a distance $r=vt$. As a result, the diffusion packet is localized in the space limited by this distance. We show below, that this localization significantly changes the shape of the diffusion packet.

\paragraph{Case $1<\alpha<2$} In this case, the mathematical expectation $p_\xi(x)$ exists. Applying Laplace transformation to (\ref{eq:racePDF}) we get $ \hat p_0(\lambda)=\alpha x_0^\alpha\int_{x_0}^{\infty}x^{-\alpha-1}e^{-\lambda x}dx$.
Integrating twice this equation in parts we obtain:
$$
\hat p_0(\lambda)=e^{-\lambda x_0}+\frac{\lambda x_0}{1-\alpha}e^{-\lambda x_0}-\frac{(\lambda x_0)^\alpha}{1-\alpha}\Gamma(2-\alpha,\lambda x_0).
$$
To get the asymptotic function at $t\to\infty$ the Tauber's theorem should be applied again. Expanding the exponents in series and omitting the terms with a power higher than $\alpha$ we obtain:
\begin{equation}\label{eq:p0LaplaceA_ge1}
\hat p_0(\lambda)\approx 1+\frac{\alpha x_0\lambda}{1-\alpha}-\frac{(\lambda x_0)^\alpha}{1-\alpha}\Gamma(2-\alpha).
\end{equation}

Substituting (\ref{eq:p0LaplaceA_ge1}), (\ref{eq:PnUpper}) in (\ref{eq:m2laplace}) we come to the asymptotic function of the second moment:
$$
\hat m_2(\lambda)=\frac{2 x_0^\alpha(\alpha-1)\Gamma(2-\alpha)v^{2-\alpha}}{\alpha(x_0/v)\lambda^{4-\alpha}- (x_0/v)^\alpha\Gamma(2-\alpha)\lambda^3}.
$$
Since $1<\alpha<2$,  at $\lambda\to0$ the second term in the denominator is negligible in comparison with the first term:
$$
\hat m_2(\lambda)\approx\frac{2 x_0^{\alpha-1}(\alpha-1)v^{3-\alpha}}{\alpha(3-\alpha)(2-\alpha)}
\frac{\Gamma(4-\alpha)}{\lambda^{4-\alpha}}.
$$
An inverse Laplace transform converts this expression to the asymptotic at $t\to\infty$
\begin{equation}\label{eq:m2a2}
M_2^N(t)\approx\frac{2 x_0^{\alpha-1}(\alpha-1)}{\alpha(3-\alpha)(2-\alpha)}(vt)^{3-\alpha}.
\end{equation}
On the same way the expressions for other moments could be obtained. The expressions for $M_4^N(t), M_6^N(t), M_8^N(t), M_{10}^N(t)$ are listed in \ref{app:MomentsExpr} (\ref{eq:m4a2})~-~(\ref{eq:m10a2}).

One can see from the expression for $M_2^N(t)$ that in the considered case the diffusion packet $\Delta(t)$ expands proportionally to $t^{(3-\alpha)/2}$ that corresponds to superdiffusion. The packet expands more slowly than in ballistic regime and so the kinematic restriction $|x|\leqslant vt$ does not affect its shape. It is shown below that in this case the effect of a finite velocity is reduced just to replacement of the diffusion coefficient $D\to D_v$, ($D_v<D$) in the equation for superdiffusion.

\section{Analysis of moments}\label{seq:momAnalysis}

\begin{figure}
\includegraphics[width=0.4\textwidth]{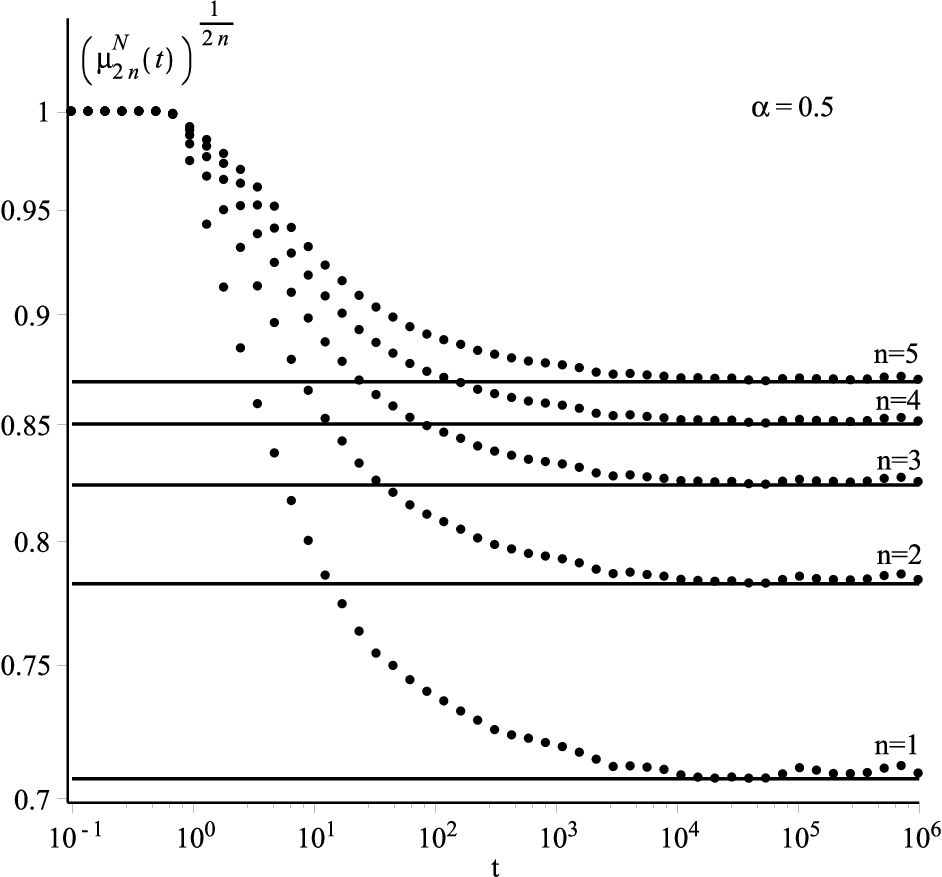}\hfill
\includegraphics[width=0.4\textwidth]{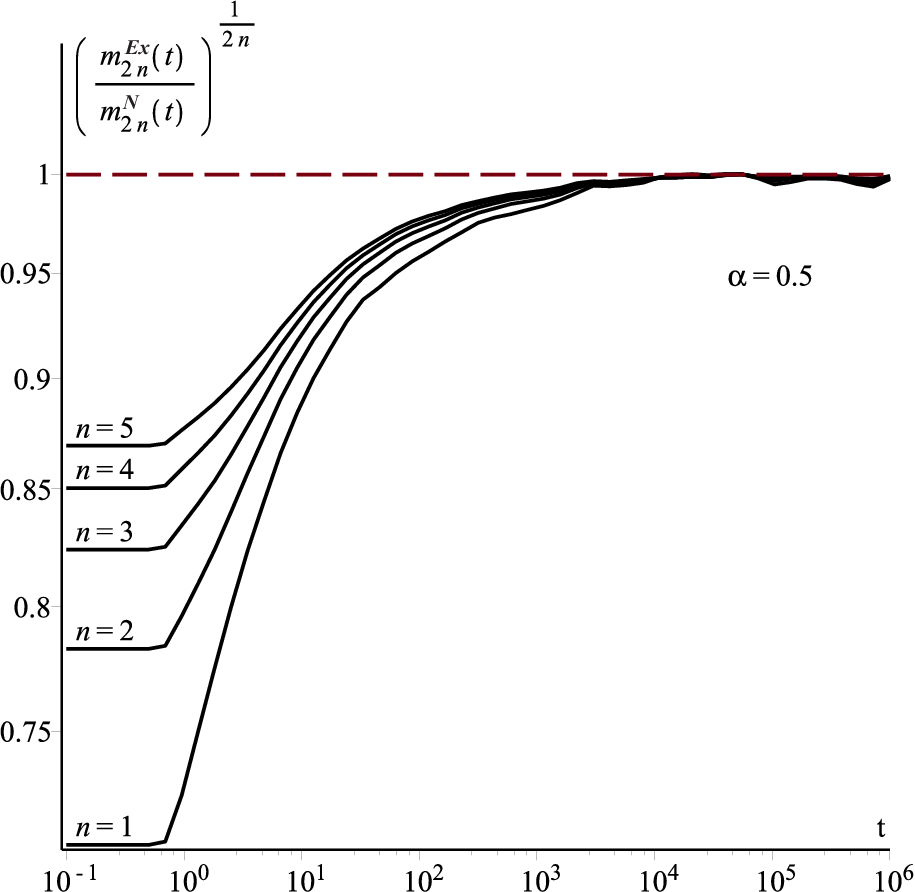}\\
\parbox[t]{0.48\textwidth}{\caption{\small Time dependences of first five even moments $\sqrt[2n]{\mu_{2n}^N(t)}$,  $n=1,2,3,4,5$ for one-dimension walk ($N=1$), the case $0<\alpha\leqslant1$  . Here, $\alpha=0.5$, $v=1$. Points are simulated by the Monte-Carlo method, solid-curves are asymptotic functions (\ref{eq:m2a01}), (\ref{eq:m4a01})~-~(\ref{eq:m10a01}).}\label{fig:mmt_m2n_n}}\hfill
\parbox[t]{0.48\textwidth}{\caption{\small Time dependencies of the exact moments $m_{2n}^{Ex}(t)$ normalized to their asymptotic functions $m_{2n}^N(t)$ at $\alpha=0.5$.}\label{fig:mmtRel_a01_d1}}
\end{figure}

\begin{figure}
\includegraphics[width=0.4\textwidth]{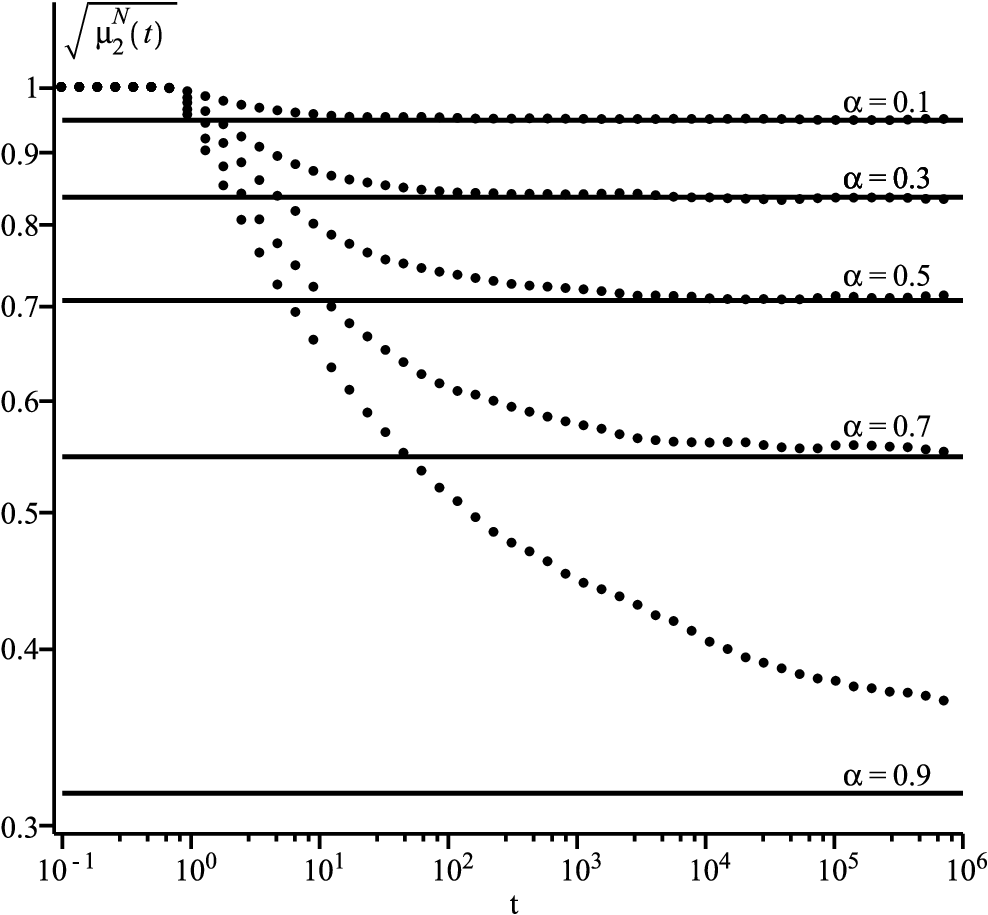}\hfill
\includegraphics[width=0.4\textwidth]{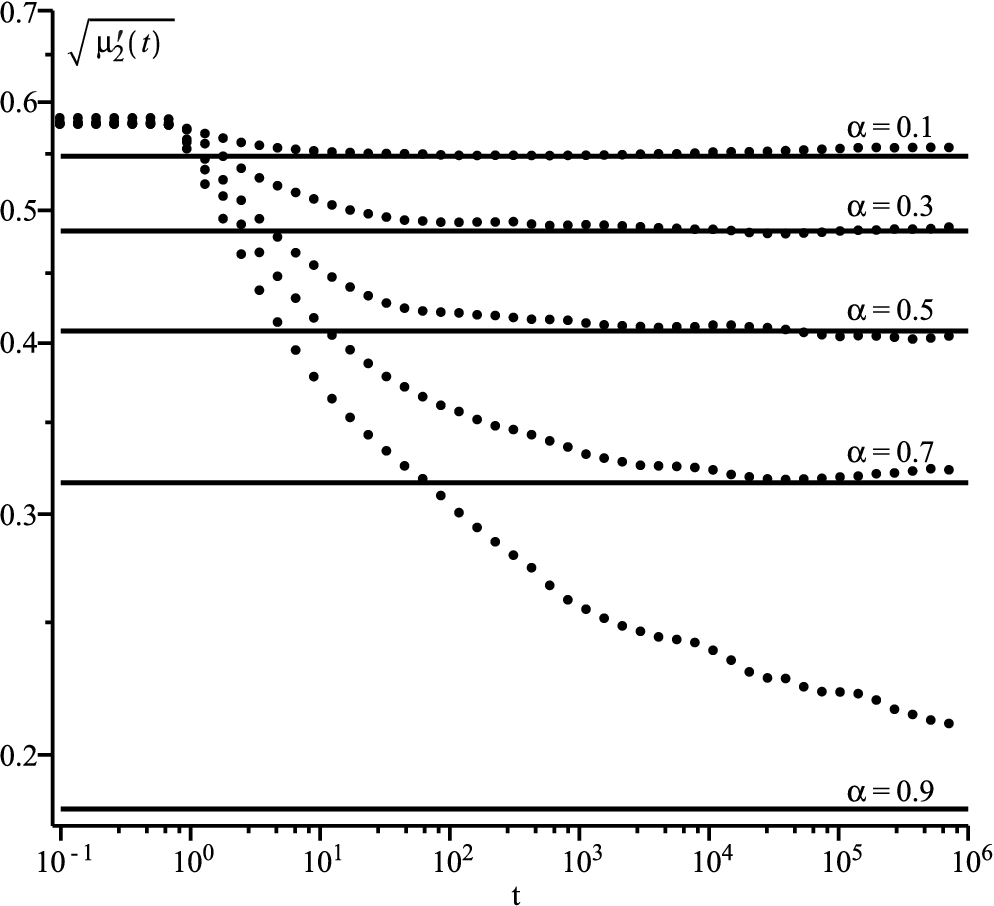}\\
\parbox[t]{0.48\textwidth}{\caption{\small Time dependences of the second moment at different $\alpha=0.1, 0.3, 0.5, 0.7, 0.9$, the case of $0<\alpha\leqslant1$ , $N = 1$ and $v=1$. Points are simulated by the Monte-Carlo method, solid-curves are asymptotic functions (\ref{eq:m2a01}).}\label{fig:mmt_m2_alpha}}\hfill
\parbox[t]{0.48\textwidth}{\caption{\small Time dependences of  $\sqrt{\mu'_{2}(t)}$ at $\alpha=0.1, 0.3, 0.5, 0.7, 0.9$ for infinite plan source and  $v=1$. Points are simulated by the Monte-Carlo method, solid-curves are asymptotic functions (\ref{eq:m2a01}).}\label{fig:mmt_m2_alpha_d3}}\hfill
\end{figure}

To simplify the analysis we consider the quantity $\mu_n^N(t)=m_n^N(t)/(vt)^n$, where $m_n^N(t)$ is the moment of order $n$. Fig.~\ref{fig:mmt_m2n_n} shows the results of calculations of $\sqrt[2n]{\mu_{2n}^N(t)}$, with $n=1,2,3,4,5$ in one-dimensional case at $\alpha=0.5$. One can see that for small times the value $\sqrt[2n]{\mu^N_{2n}(t)}$ corresponding to the exact moment is constant. It means that during short times after generation the particle propagates in a ballistic regime. It is obviously that immediately after generation the particle moves without scattering producing a straight trajectory. As result, all particles are localized on the surface $r=vt$ forming the front of the distribution. We call this regime the ballistic regime. For the latter times the scattering processes become important resulting in formation of the pre-asymptotic distribution. This propagation mode is called the diffusion mode. As can be seen from the calculations, the transition from ballistic to diffusion regime happens at the time moment $t\approx1$.  Then the diffusion regime transforms into the asymptotic regime. It happens when the exact moment coincides with the asymptotics. Figure~\ref{fig:mmtRel_a01_d1} shows the exact moments $m_{2n}^{Ex}(t)$ normalized to their asymptotics $m_{2n}^N(t)$. One can see that at the given $\alpha$ the moments of all order get asymptotic behavior at nearly the same time. So, the moments converge to their asymptotics uniformly. However, the time when the moments achieve their asymptotics depends on $\alpha$ (Fig.~\ref{fig:mmt_m2_alpha}). We will use this time parameter $T^*$ in our consideration. From Fig.~\ref{fig:mmt_m2_alpha} one can determine that at $\alpha=0.1$ the time $T^*\approx10$, at $\alpha=0.3$ $T^*\approx100$, at $\alpha=0.5$ $T^*\approx10^4$, at $\alpha=0.7$ $T^*\approx10^5$, and at $\alpha=0.9$ the time $T^*$ significantly exceeds $10^6$.

There is one more feature of these asymptotic distributions. For the considered case the asymptotic behavior is characterized by the dependence $m_{2n}^N(t)\propto(vt)^{2n}$. In Figs. ~\ref{fig:mmt_m2n_n}~and~\ref{fig:mmt_m2_alpha}, it transforms into a constant value. This behavior characterizes quasi-ballistic regime of the package expansion. Hence, at $0<\alpha<1$ the kinematic restriction keeps its dominant influence on the formation of the diffusion packet even in the asymptotic regime.

To analyze three-dimensional case we simplify the problem. Let us consider the distribution of the $x$-component $X(t)$ of the random vector $\mathbf{R}_N(t)$. The moments of the radius-vector describing a particle walking in $N$-dimensional isotropic media with a point source have been already found. From the transport theory the density distribution of $X(t)$ is the density of an infinite isotropic plane source with a unit surface density. A relation between the moments considered in these two problems (with point and plane isotropic sources) is the same as in the stationary case \cite{Case1976}
\begin{equation}\label{eq:mmt_sl_pt}
m_{2n}'(t)=\langle X^{2n}(t)\rangle=\langle R^{2n}_N(t) \cos^{2n}\theta\rangle=m^N_{2n}(t)/(2n+1),
\end{equation}
where $m_{2n}'(t)$ is the moment for a plane source. It worth nothing that this relation is correct for $N = 3$ only.

The calculation results for the second moment is shown in Figure~\ref{fig:mmt_m2_alpha_d3}, where $\mu_2'(t)=m_2'(t)/(vt)^2$. One can see that the behaviors of the moments in three and one dimensional cases are similar. There are still three modes: ballistic, diffusion and asymptotic. The exact moments converge uniformly to its asymptotic behavior for the time that depends on $\alpha$: the larger $\alpha$ the longer the time. Moments increase as $m_{2n}'(t)\propto(vt)^{2n}$, and so the kinematic restriction has a dominating influence on the formation of the diffusion package in three-dimensional case. Times to get the asymptotic behavior are $T^*=10,100,2000,2\cdot10^4$  and for $\alpha=0.9$ the time is much longer than $10^6$.

\begin{figure}
\includegraphics[width=0.4\textwidth]{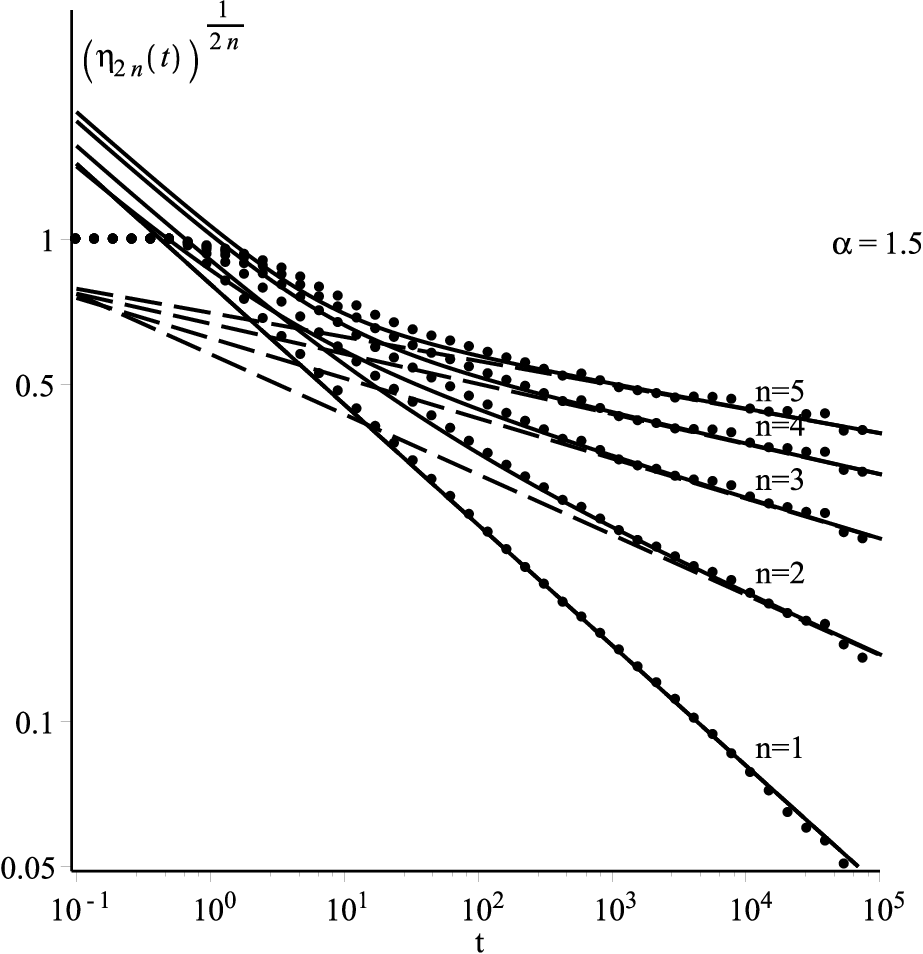}\hfill
\includegraphics[width=0.4\textwidth]{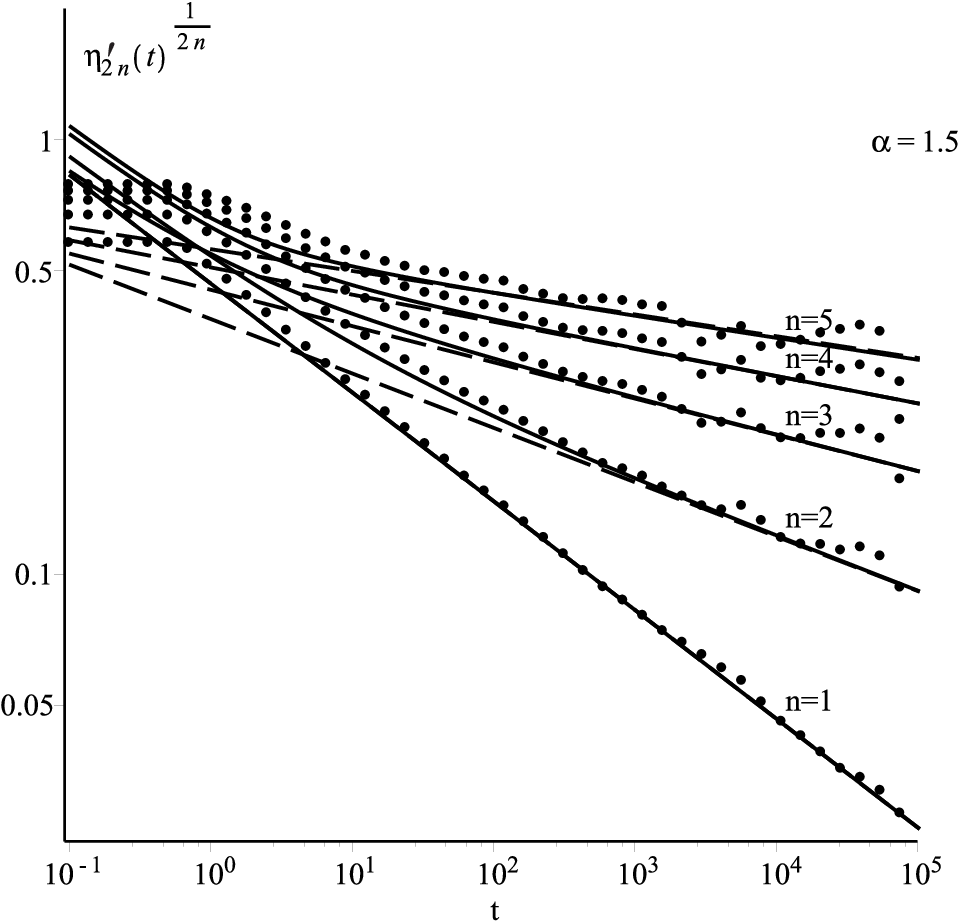}\\
\parbox[t]{0.48\textwidth}{\caption{\small Time dependences of $\sqrt[2n]{\eta_{2n}(t)}$ for one-dimensional case, at $\alpha=1.5$ and different $n$. Points are simulated by Monte-Carlo method, solid-curves are asymptotic functions (\ref{eq:m2a2}), (\ref{eq:m4a2})~-~(\ref{eq:m10a2}), dashed curves are the main terms of the asymptotic functions.}\label{fig:mmt_m2n_a12_n_d1}}\hfill
\parbox[t]{0.48\textwidth}{\caption{\small Time dependences of $\sqrt[2n]{\eta'_{2n}(t)}$ for three-dimensional walk, at $\alpha=1.5$ and different $n$. Points are simulated by the Monte-Carlo method, solid-curves are asymptotic functions (\ref{eq:m2a2}), (\ref{eq:m4a2})~-~(\ref{eq:m10a2}), dashed curves are the main terms of the asymptotic functions.
}\label{fig:mmt_m2n_a12_n_d3}}
\end{figure}

In the case of $1<\alpha<2$  it is suitable to use $\eta_{2n}(t)=M_{2n}^N(t)/(vt)^{2n}$. The results for the one-and three- dimensional cases are shown in Fig.~\ref{fig:mmt_m2n_a12_n_d1}~and~\ref{fig:mmt_m2n_a12_n_d3}, respectively. One can see that the process regimes are divided into ballistic, diffusive, and asymptotic regimes. Ballistic regime is realized at times $t\leqslant1$.

\begin{figure}
\includegraphics[width=0.485\textwidth]{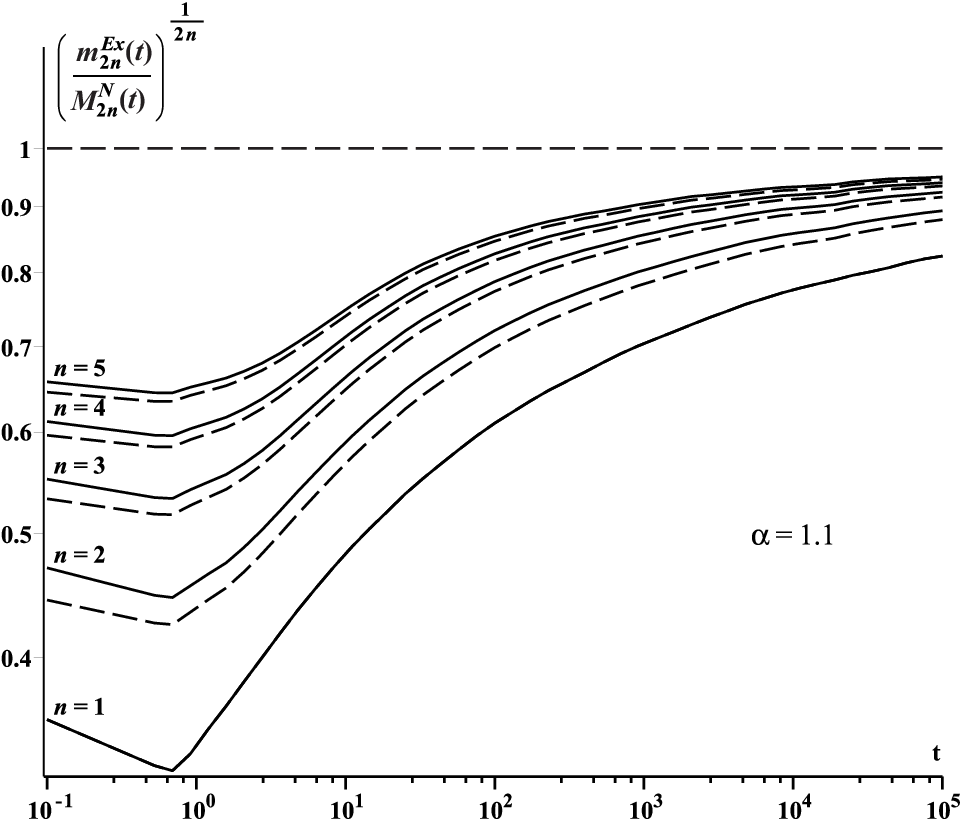}\hfill
\includegraphics[width=0.485\textwidth]{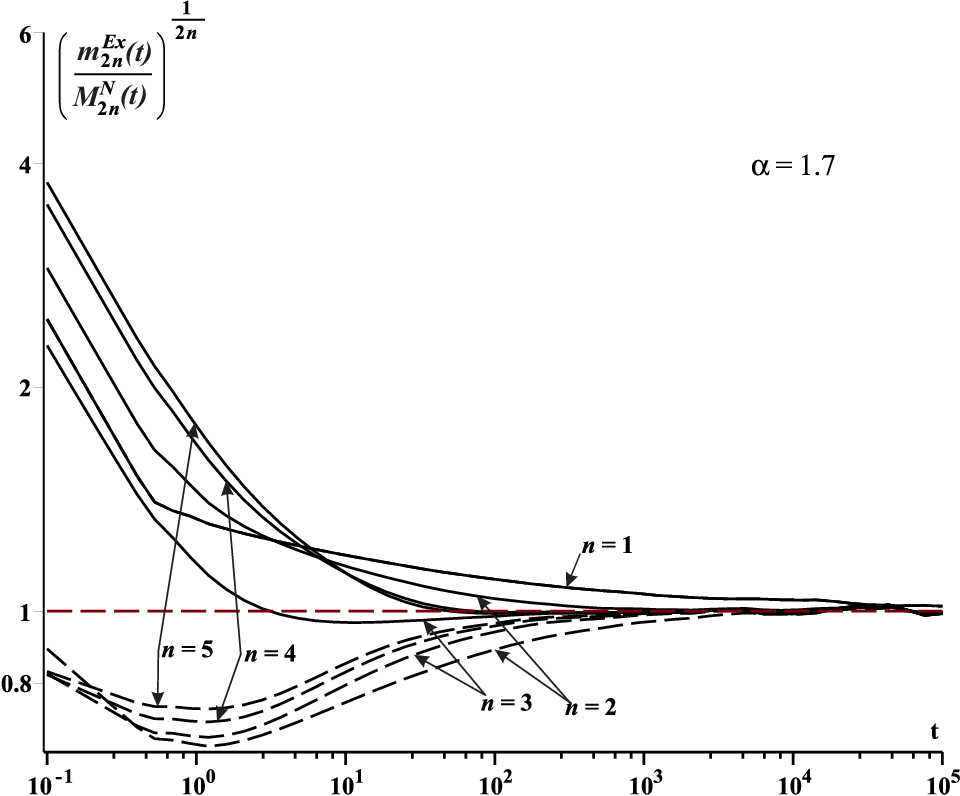}\\
\parbox[t]{0.99\textwidth}{\caption{\small Time dependences of the exact moments $m_{2n}^{Ex}(t)$ normalized to their asymptotic functions $M_{2n}^N(t)$ at $\alpha=1.1$ (left) and $\alpha=1.7$ (right). Solid curves are asymptotic functions (\ref{eq:m2a2}), (\ref{eq:m4a2})~-~(\ref{eq:m10a2}), dashed curves are the main terms of the asymptotic functions.}\label{fig:mmtRel_m2n}}
\end{figure}

Let us investigate the time moment the process gets asymptotic behavior. As shown in \cite{JarovikovaPhD2001_en} the exact moments converge to their asymptotics nonuniformly. It means that you can not specify the moment of time when all obtained asymptotics differ from the exact moments by a small arbitrate value. A similar situation is observed in our case. Fig.~\ref{fig:mmtRel_m2n} shows the ratio $\sqrt[2n]{m_{2n}^{Ex}(t)/M_{2n}^N(t)}$. One can see that the exact moments achieve their main asymptotics in different times. Accounting the higher asymptotic terms improves the situation. Moments $M_{2n}^N(t)$ (solid curves) converge to the exact values faster than the moments with the main asymptotic term only (dashed curves). The rule is the higher the order of the moment, the faster it converges to the asymptotic behavior. It means that the second order exact moment is getting the asymptotic behavior most slowly. Therefore, to determine the time $T^*$ in the case $1<\alpha<2$  the second-order moment should be considered.

\section{Reconstruction of the particles distribution}\label{sec:PDFReconstr}

The method of moments allows to restore the asymptotic distribution density. The idea of this method is to restore the density using a system of orthogonal polynomials. Such system of polynomials should be taken so the weight function of the system most accurately fits the shape of the distribution. In one-dimensional case, simulation based on the Monte Carlo method demonstrates that for $0<\alpha<1$  and $1<\alpha<2$ the distributions are $U$-shaped and bell-shaped, respectively. Based on this information the system of Chebyshev polynomials $T_n(x)$ suits distribution recovery in the first case, while Hermite polynomials $H_{n}(x)$ has to be employed in the second.

\paragraph{Case $0<\alpha<1$} First, we consider one-dimensional random walk with a point source. To reconstruct the distribution we use a system of Chebyshev orthogonal polynomials of the 1-st kind $T_n(x)$ with a weight function $w(\xi)=1/\sqrt{1-\xi^2}$. Since the Chebyshev polynomials are orthogonal in the interval $[- 1,1]$ for the reconstruction of the distribution density we have to transform coordinates in such a way that all density is located in this segment. For particle propagating with a finite velocity, the distribution density is concentrated within the interval $-vt\leqslant x\leqslant vt$.  Outside this interval $p(x,t)=0$, so change of variable $-1\leqslant\xi\leqslant1$, where $\xi=x/vt$ allows to employ Chebyshev polynomials. For this transition the distribution density and moments are transformed as $p(\xi,t)=p(vt\xi,t)/(vt), \mu_n^N(t)=m_n^N(t)/(vt)^n$, where $m_n^N(t)$ are the moments of the distribution $p(x,t)$ and  $\mu_n^N(t)$ the moments of the distribution $p(\xi,t)$.

Using expressions for Chebyshev polynomials we   come to expansion of the distribution density:
\begin{equation}\label{eq:pdf_a1}
p(\xi,t)\approx\frac{1}{\sqrt{1-\xi^2}}\sum_{k=0}^5c_{2k}(t)T_{2k}(\xi), \quad t>T^*,
\end{equation}
where the coefficients are:
$$
c_l(t)=\frac{l}{2h_l}\sum_{m=0}^{[l/2]}\frac{(-1)^m(l-m-1)!2^{l-2m}}{m!(l-2m)!}\mu_{l-2m}^N(t).
$$
Here $[A]$ is an integer part of $A$. Since asymptotic functions of the moments have been used for the reconstructing, the expansion (\ref{eq:pdf_a1}) is valid for asymptotic regime only. In previous section we found that the time the moments achieve the asymptotic behaviors increases with the increase of the parameter $\alpha$. In particular, at $\alpha=0.7$ the time $T^*\approx 10^5$, and at $\alpha=0.9$  the time exceeds $10^6$ . To be concrete we put $T^*=10^4$ in our calculations.

\begin{figure}
\includegraphics[width=0.45\textwidth]{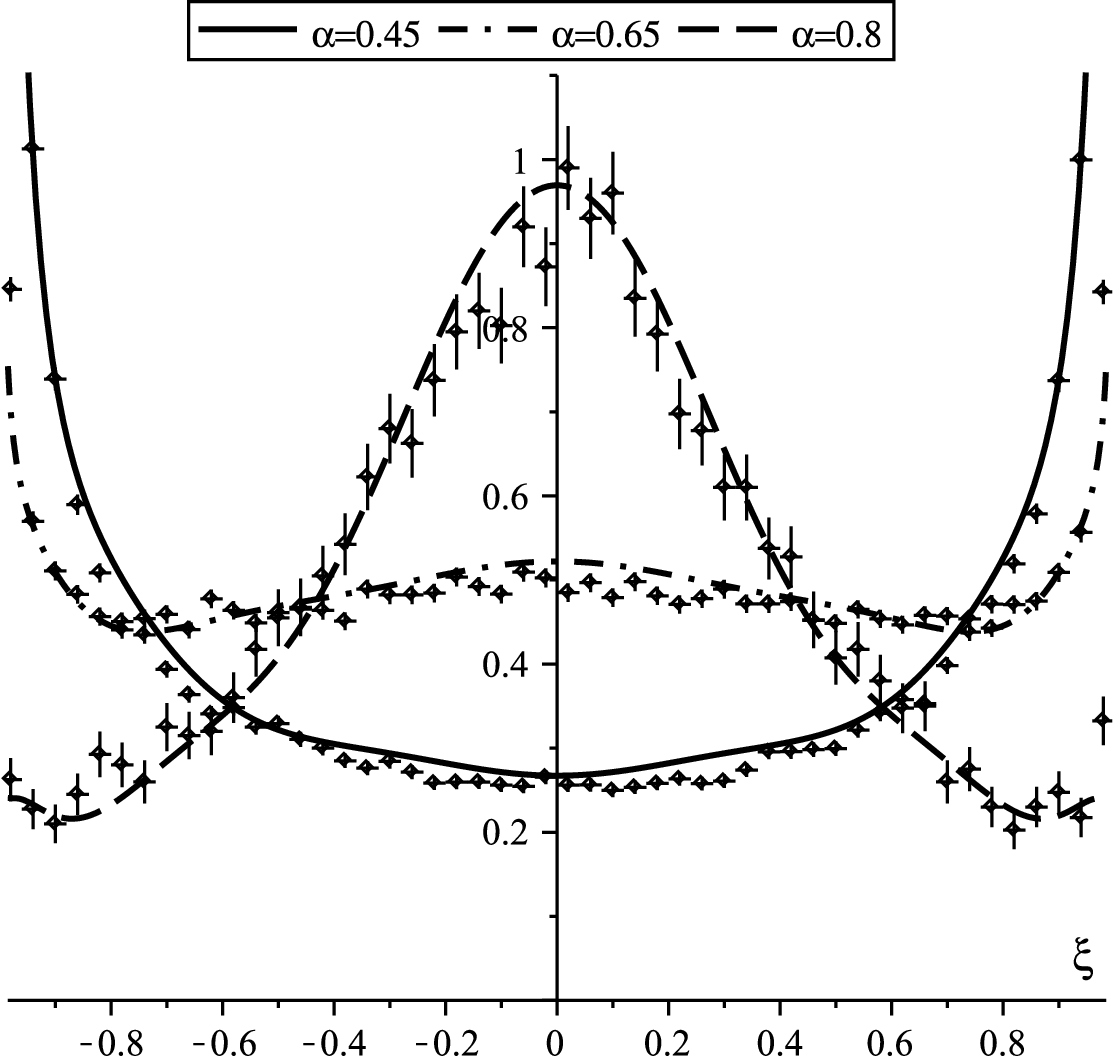}\hfill
\includegraphics[width=0.45\textwidth]{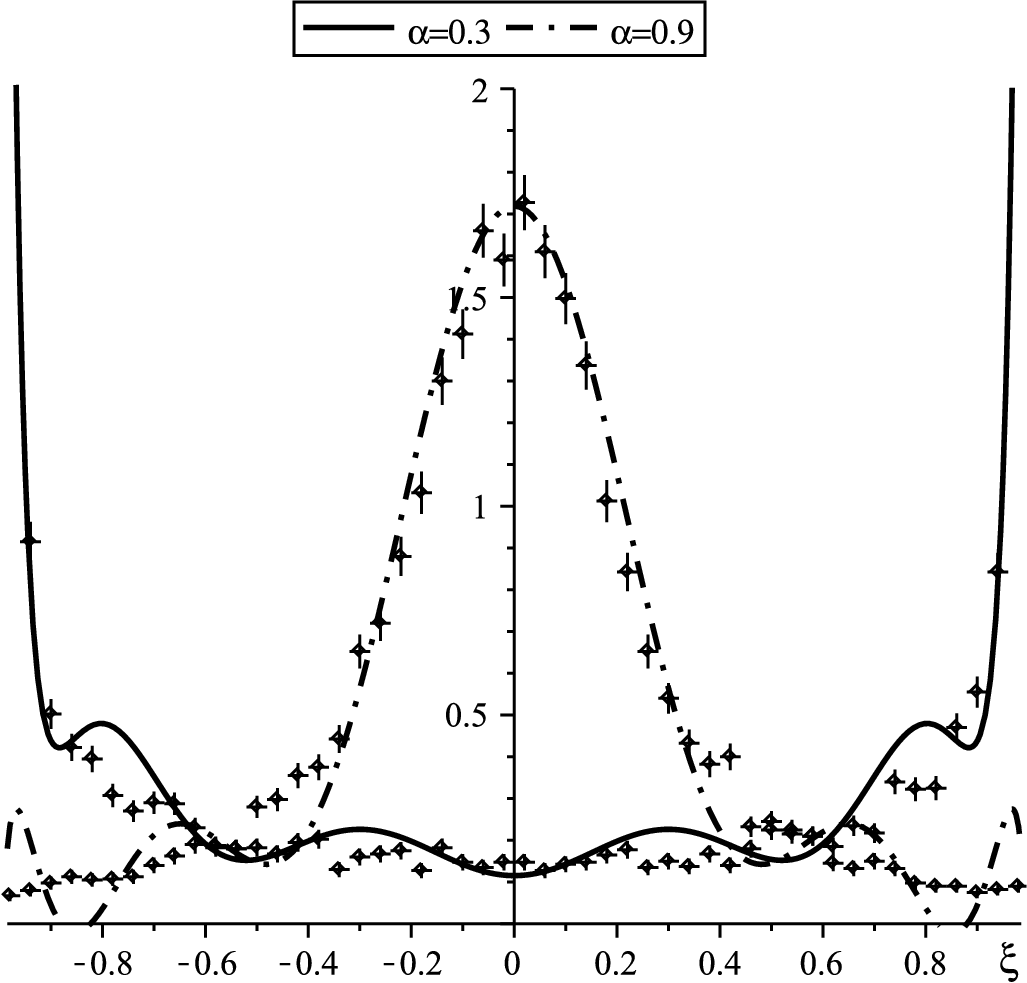}\\
\caption{\small The reconstructed distribution density for the case  of  $0<\alpha<1$ at different $\alpha$, $N=1$, $v=1$, $t=10000$. Points are simulated by the Monte-Carlo method, curves are asymptotic functions (\ref{eq:pdf_a1}).\label{fig:pdf1}}
\end{figure}

The results are shown in Fig.~\ref{fig:pdf1}. One can see from Fig.~\ref{fig:pdf1} (left) that for $0.4\leqslant\alpha\leqslant0.8$ the method reconstructs the distribution density with high accuracy, while for $\alpha\notin[0.4,0.8]$ it does not work well. It is explained by limited number of moments used for the reconstruction, besides not all of them have achieved their asymptotics.

Now we consider the three-dimensional case. For simplicity, the distribution of the $x$-component of the radius vector $\mathbf{R}_N(t)$ of the particle is considered. As mentioned above, the distribution density of $x$-coordinate of the particle is the distribution density of the infinite plan isotropic source. Moments of the distribution density in such source are connected with the obtained higher-order moments through the ratio (\ref{eq:mmt_sl_pt}).

For the reconstruction of the distribution density we again use the Chebyshev polynomials of the first kind. Taking into account all mentioned above, the distribution density is
\begin{equation}\label{eq:pdf_a1_d3}
p(\xi,t)\approx\frac{1}{\sqrt{1-\xi^2}}\sum_{k=0}^5c_{2k}(t)T_{2k}(\xi), \quad t\geqslant T^*,
\end{equation}
$$
c_l(t)=\frac{l}{2h_l}\sum_{m=0}^{[l/2]}\frac{(-1)^m(l-m-1)!2^{l-2m}}{m!(l-2m)!}\mu_{l-2m}'(t),
$$
where $\mu_{k}'(t)=m_k'(t)/(vt)^k$.

\begin{figure}
\includegraphics[width=0.45\textwidth]{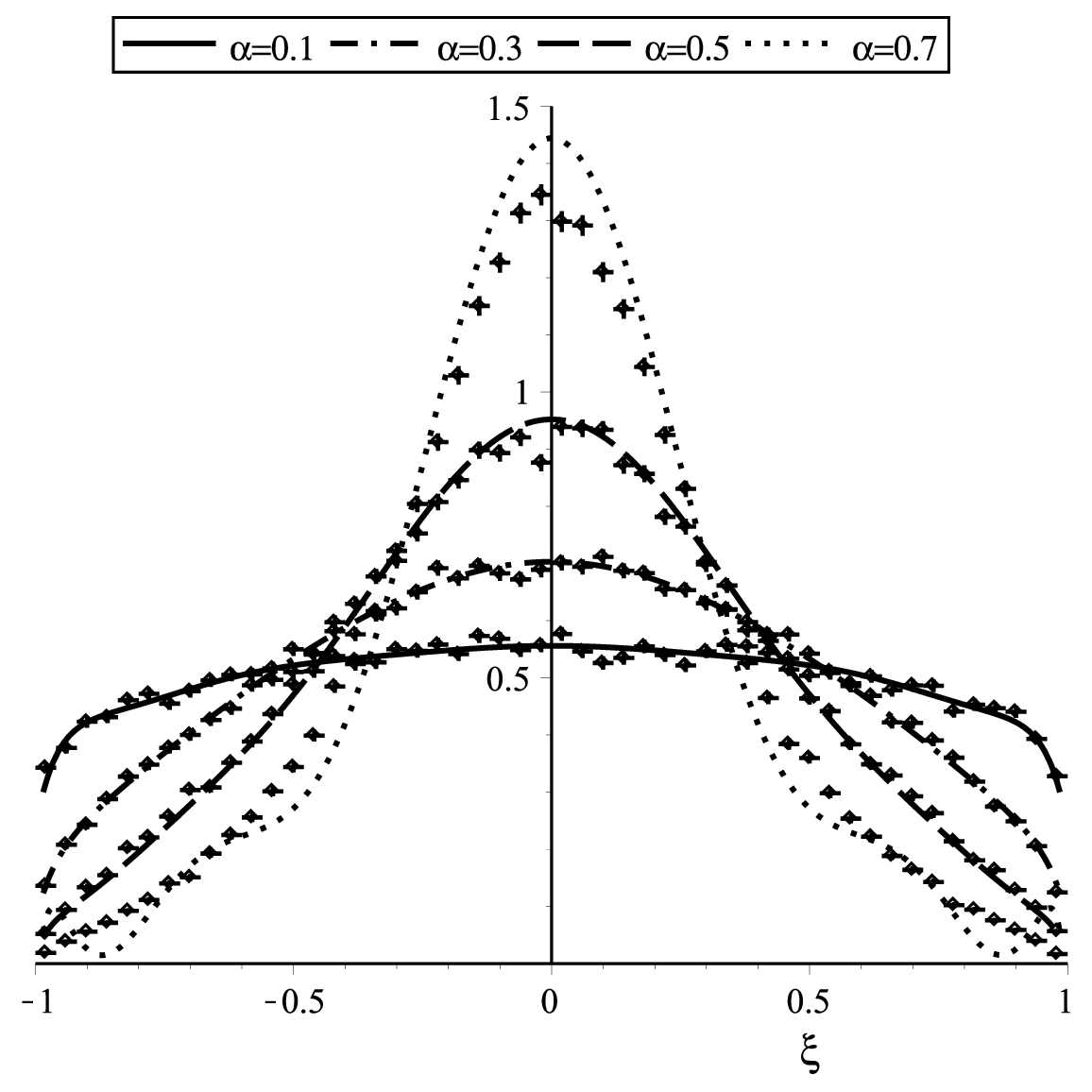}\hfill
\parbox[b]{0.48\textwidth}{\caption{\small The reconstructed distribution density for case of a plane infinitive isotropic source, $0<\alpha<1$ at  $v=1$ and $T^*=10^4$.  Points are results of simulations by the Monte-Carlo method, curves are restored by the Eq.~(\ref{eq:pdf_a1_d3}).}\label{fig:pdf_a01_d3}}
\end{figure}

Fig.~\ref{fig:pdf_a01_d3} shows the reconstruction of the distribution density (\ref{eq:pdf_a1_d3}) for the time $T^*=10^4$. At $\alpha\leqslant 0.5$ the reconstruction results coincide with the results obtained by the Monte Carlo method. For $\alpha=0.7$ a limiting number of the moments in the density expansion restricts the reconstruction.

\paragraph{Case $1<\alpha<2$} In this case, a mathematical expectation of the particle path exists and, at $\alpha=2$  their dispersion exists also. At $\alpha=2$ the process gets the normal diffusion regime with the normal distribution in limiting case. Therefore, for reconstruction of the distribution employing Hermite polynomials $H_{n}(x)$ should be employed. For better recovery the weight function must be redefined for best fit of the recovered distribution. Therefore, the weight function with the Hermite polynomials is chosen as $w(x)=\exp\left(-x^2/2\sigma^2\right)/\sqrt{2\pi\sigma^2}$, where $\sigma^2=M_2^N(t)$, resulting in the following Hermite polynomials expressions:
$$
H_{n}(x)=\frac{n!}{\sigma^{2n}}\sum_{m=0}^{[n/2]}\frac{(-1)^m\sigma^{2m}}{m!2^m(n-2m)!}x^{n-2m},
$$
with $h_k=k!/\sigma^{2k}$ as a condition of their orthogonality.

\begin{figure}
\includegraphics[width=0.45\textwidth]{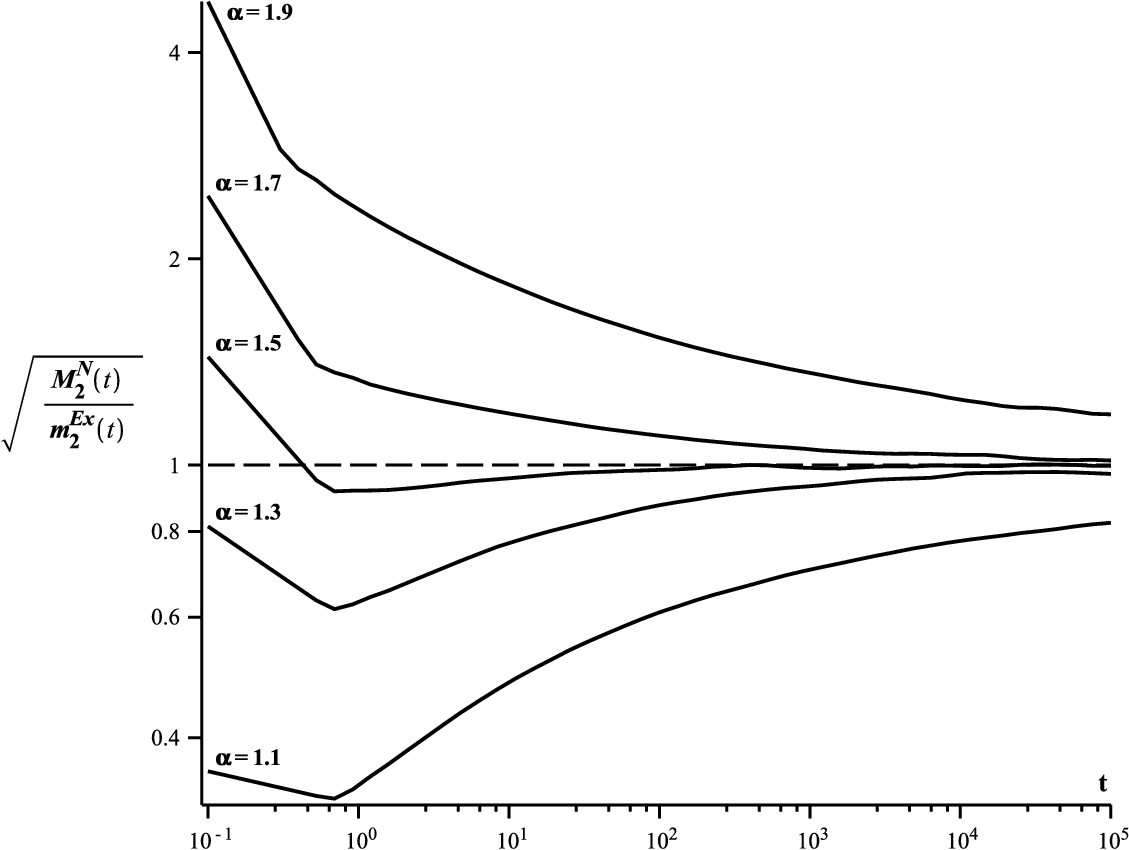}\hfill
\includegraphics[width=0.45\textwidth]{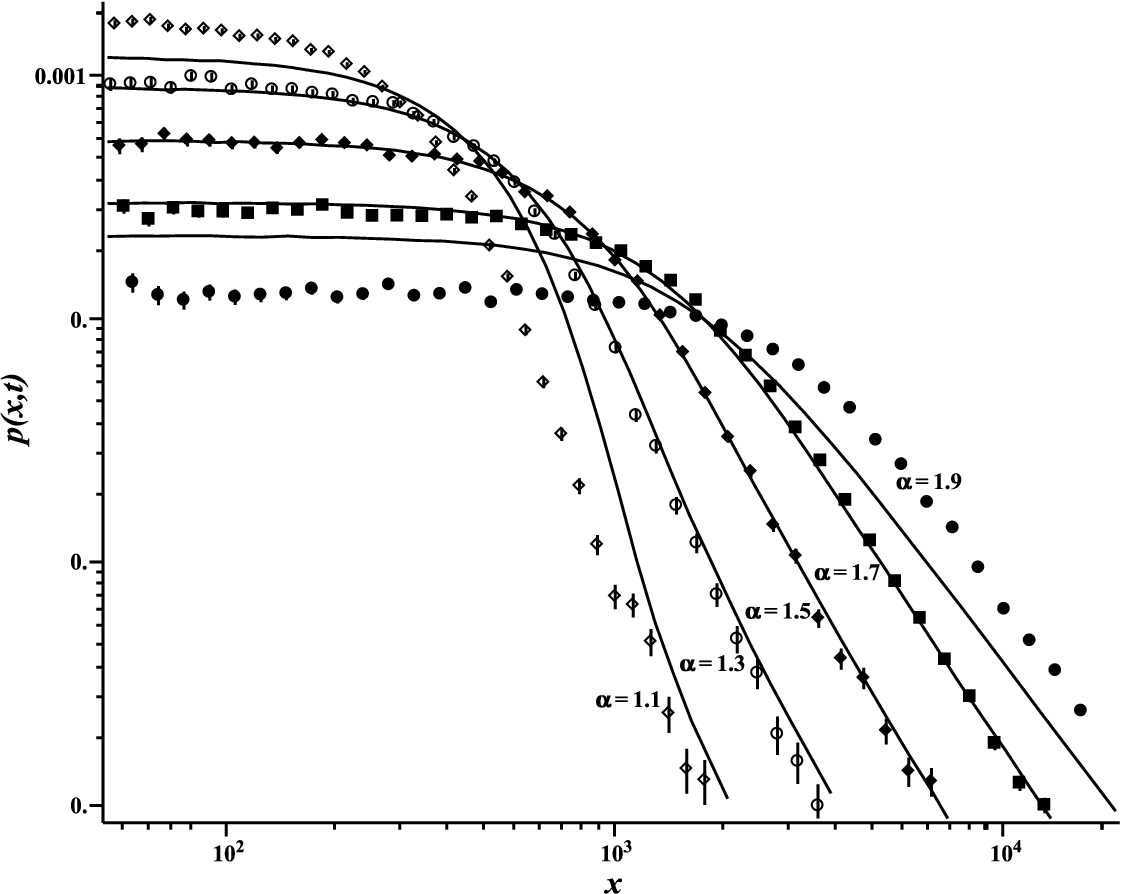}\\
\parbox[t]{0.48\textwidth}{\caption{\small Time dependence of the ratio between asymptiotics of the second moment $M_2^N(t)$ and its exact value $m^{Ex}_2(t)$ for $1<\alpha<2$.}\label{fig:mmtRel_m2a11_19}}\hfill
\parbox[t]{0.48\textwidth}{\caption{\small Asymptotic distribution of particles for $1<\alpha<2$ at $t=10^4$. Points are the results obtained by the Monte Carlo method, solid curves are solution of the fractional diffusion equation with  $D_v$.}\label{fig:pdf_a11_19}}
\end{figure}

Nevertheless, the reconstruction of the distribution with Hermite polynomials was not successful. The reasons of that is a small number of the moments used for the reconstruction and their non-uniform and slow convergence to their asymptotics (see. Fig.~\ref{fig:mmtRel_m2a11_19}). It can be seen that at $\alpha=1.1$ and $1.9$ the asymptotic behavior is achieved at$t\geqslant 10^5$, $\alpha=1.3$ and $1.7$ for values of $t\approx10^5$, and at $\alpha=1.5$ for 20. It should be noted that at the given $\alpha$ high-order moments achieve their asymptotics faster.

However, in this case, we could obtain the asymptotic distribution of the particles as well. Indeed, the kinematic restriction does not affect the formation of the asymptotic distribution. Therefore, this distribution is expressed in terms of fractional diffusion equation (see.~\cite{Zolotarev1999})
\begin{equation}\label{eq:supDfsEq}
\frac{\partial p(x,t)}{\partial t}=-D(-\Delta)^{\alpha/2}p(x,t).
\end{equation}
A solution of this equation has a form $ p(x,t)=\left(Dt\right)^{-N/\alpha} g^{(\alpha)}\left(x\left(Dt\right)^{-1/\alpha}\right), \alpha>1$, where $g^{(\alpha)}(x)$ is the $N$-dimentional Levy stable law, $D$ is the diffusion coefficient. Note that a Levy walk with finite velocity and traps has been considered in  \cite{Zolotarev1999}. It has been reported that accounting of the finite velocity leads to slowing of the process of diffusion package expansion in comparison with the case of  $v=\infty$. This slowing could be taken into account by replacement of the diffusion coefficient $D\to D_v$ in the Eq.~(\ref{eq:supDfsEq}). In the case of a Levy walk without traps the expression for the diffusion coefficient is reduced to $D_v=(v/a)D$, 	where $a$  is an average free path of the particle. For the density (\ref{eq:racePDF}) we can get $a=\mean\xi=\tfrac{\alpha}{\alpha-1}x_0$. Taking above mentioned into account, the Eq.~(\ref{eq:supDfsEq}) and its solution could be written as
$$
\frac{\partial p(x,t)}{\partial t}=-D_v(-\Delta)^{\alpha/2}p(x,t),
$$
\begin{equation}\label{eq:solSupDfsDv}
p(x,t)=\left(D_vt\right)^{-N/\alpha} g^{(\alpha)}\left(x\left(D_vt\right)^{-1/\alpha}\right), \alpha>1,
\end{equation}
where $D_v=\frac{v(\alpha-1)}{\alpha x_0}D$.

The solution (\ref{eq:solSupDfsDv}) in one-dimensional case ($N=1$) at different $\alpha$ (solid curves) and the solution obtained by the Monte-Carlo methods for time $t=10^4$ are shown   in Fig.~\ref{fig:pdf_a11_19}. One can see that at $\alpha=1.3, 1.5, 1.7$ a good agreement between asymptotic and exact solutions is achieved. At $\alpha=1.1, 1.9$ the distribution (\ref{eq:solSupDfsDv}) and the exact solution of the kinetic equation are not in total agreement. It is explained by the fact that at time $t=10^4$ the walk process does not get its asymptotic. Indeed, at time $t=10^4$ and $\alpha=1.1, 1.9$ the asymptotic is not reached (see Fig.~\ref{fig:mmtRel_m2a11_19}), but at $\alpha=1.3, 1.5, 1.7$ the process is already in the asymptotic regime.

\section{Anomalous diffusion coefficient }\label{sec:Koeff}

The anomalous diffusion coefficient can be obtained from the relations for the second moments. Indeed, according to \cite{Metzler2000}, the anomalous diffusion coefficient is determined by relation $\langle x^2(t)\rangle \sim D_\alpha t^\mu$. Hence, from (\ref{eq:m2a01}) and (\ref{eq:m2a2}) we derive
\begin{equation}\label{eq:D_alpha}
D_\alpha=\left\{\begin{array}{ll}
(1-\alpha)v^2,& 0<\alpha<1\\
\displaystyle \frac{2 x_0^{\alpha-1}(\alpha-1)}{\alpha(3-\alpha)(2-\alpha)}v^{3-\alpha},&1<\alpha<2.
\end{array}\right.
\end{equation}
One can see that introduction of the finite velocity leads to dependence of the diffusion coefficient on the velocity. As expected, with rising velocity the anomalous diffusion coefficient increases. In the case $0<\alpha<1$, the anomalous diffusion coefficient $D_\alpha$ demonstrates quadratic dependence on the velocity. The same result has been reported in Refs.~\cite{Uchaikin1998, Uchaikin_TMF1998b_eng, Froemberg2013a}. In the case $1<\alpha<2$, $D_\alpha\sim v^{3-\alpha}$. This result is in good agreement with the results reported in Refs.~\cite{Dhar2013, Froemberg2013a, Zimbardo2013}.

However, it should be noted that in Ref.~\cite{Zimbardo2013} some inaccuracy has been made in deriving the expression for the anomalous diffusion coefficient. Here, the anomalous diffusion coefficient $D_\mu$ is determined from relation $
\langle x^2(s)\rangle=\frac{2 Av^2t_0^\mu}{\tau}\frac{\Gamma(3-\mu)}{\mu-1}s^{\mu-5}\equiv 2D_\mu s^{\mu-5}
$ (see Eq. (13) in Ref.~\cite{Zimbardo2013}). This expression itself is the Laplace transformation in time from the second moment of distribution. Thus, to obtain a correct relation for the anomalous diffusion coefficient the inverse Laplace transformation of this expression has to be performed.

The relation for asymptotic distribution of the second moments from Ref.~\cite{Zimbardo2013} coincides with Eq.~(\ref{eq:m2a2}) if the time distribution in the state of motion $\psi(t)$ (see Eq.~(15) in Ref.~\cite{Zimbardo2013}) is replaced with the following distribution :
$$
\psi(t)=\left\{\begin{array}{ll}
0,&t< t_0\\
A(t/t_0)^{-\mu},&t\geqslant t_0\end{array}\right.,
$$
that corresponds to distribution (\ref{eq:racePDF}). In this case, relation for diffusion coefficient derived in \cite{Zimbardo2013} coincides with (\ref{eq:D_alpha}) for the case $1<\alpha<2$.

Refs.~\cite{Dhar2013, Schmiedeberg2009, Froemberg2013a}, that are also important for our study, report expressions for the second moments for the considered process that correspond to (\ref{eq:m2a01}) and (\ref{eq:m2a2}). In Ref.~\cite{Dhar2013}, the authors obtain relation for the fourth moment for the case $1<\alpha<2$ that completely coincides with the main asymptotic term in Eq.~(\ref{eq:m4a2}). Asymptotic behavior of the second moment agrees with the results reported in Refs.~\cite{KlafterJ1994, Froemberg2013a}.

\section{Discussion}\label{sec:disscusion}

As mentioned in the introduction accounting the finite speed with anomalous diffusion has been already studied. Summarizing the results, one can say that for the anomalous diffusion with an infinite expectation of the free-path accounting of the finite velocity leads to replacement of the fractional Laplacian in the fractional diffusion equation by the material derivative of fractional order \cite{Zaburdaev2002,Sokolov2003,Uchaikin2011,Uchaikin2011a}. The limit distributions in one-dimensional case are $U$-shaped and $W$-shaped \cite{Zolotarev1999, Froemberg2015,JarovikovaPhD2001_en,Uchaikin2003a,Zaburdaev2008, Uchaikin:JETPL:2010, Uchaikin2011, Uchaikin2011a, Uchaikin2014} and described by the Lamperty distribution \cite{Uchaikin2011,Uchaikin2011a,Uchaikin2009,Froemberg2013a}.  However, all the reported results are for one-dimensional case and the reasons responsible for formation of such kinds of distributions are not discussed.

In this paper we consider a multi-dimensional random Levy walk with finite velocity without traps. We use a distribution of free paths of a power-law and study the asymptotic distribution of particles employing the method of moments. The recurrent equation (\ref{eq:m2n}) for the moment of any order $2n, n=1,2,3,\dots$ of $N$-dimensional vector $\mathbf{R}_N(t)$ describing the position of the particle is derived. The study of asymptotics of the moments leads us to the consideration of two cases: 1) $0<\alpha<1$ (the mathematical expectation and variance of free paths are infinite); 2) $1<\alpha<2$ (the mathematical expectation of the free path distribution exists, but the variance is infinite).

In the first case, the expression of the second moment (\ref{eq:m2a01}) shows that it does not depend on the dimension and the width of the diffusion packet grows as $\Delta(t)\equiv\sqrt{m_2^N(t)}\propto t$. This result is in agreement with the results of other authors \cite{Schmiedeberg2009,Zumofen1993,Klafter1993, KlafterJ1994, Uchaikin_TMF1998b_eng, Uchaikin2003a, Froemberg2013a}. Such expansion corresponds to the quasi-ballistic regime of superdiffussion that is characterized by the expansion proportional to the velocity of free particles. So in the case of finite speed the dependence on $\alpha$  of width diffusion packet expansion disappears. Indeed, according to the CTRW model the diffusion packet expands as $\Delta(t)\propto t^{\gamma(\alpha)}$, where $\gamma(\alpha)=1/\alpha$, while in the considered case $\gamma=1$ for the whole range $0<\alpha<1$. From (\ref{eq:m2a01}) the rate of the diffusion packet expansion $v_{\Delta(t)}=\sqrt{1-\alpha}v$, $v_{\Delta(t)}\to v$ at $\alpha\to0$ and $v_{\Delta(t)}\to0$ at $\alpha\to1$. In one-dimensional case this forms two different shapes of the diffusion package: $U$-shaped, with $\alpha\lesssim0.6$ and $W$-shaped, with $\alpha>0.6$ (see Fig.~\ref{fig:pdf1})).

The reason of $U$- and $W$- shape formation is the effect of kinematic restriction $|x|\leqslant vt$. Since the velocity is finite at time moment $t$ the particle cannot leave the area $|x|>vt$.  It is localized in this area.  Since in this case the distribution (\ref{eq:racePDF}) has an infinite expectation, the distribution of this random value has a long tail, and the lower $\alpha$ the higher the probability concentrated in the tail. Being localized by kinematic restriction $|x|\leqslant vt$ the particle cannot leave this area. Since a significant part of the probability is in the tail, more and more particles are concentrated in area neighboring the lines $x=vt$  and $x=-vt$ as the parameter $\alpha$ decreases. It causes formation of $U$- and $W$- shape of the asymptotic distribution in one-dimensional case.

The same conclusion has been done by other authors \cite{Zolotarev1999,JarovikovaPhD2001_en,Uchaikin2003a} considering anomalous diffusion with traps. They concluded \cite{JarovikovaPhD2001_en} that the cause of $W$-shaped distribution formation is the domination of the process of capturing the particles by traps over the process of free movement. However, this conclusion is not entirely true, since such distribution could be formed in the absence of traps. Hence, the cause of particle concentration in the vicinity of zero (i.e. the formation of $W$-shaped distribution) is the scattering process, which in conjugation with the decreasing probability of long paths (with increasing $\alpha$) leads to trapping of particles near zero. Accounting of traps leads to the decrease of the diffusion coefficient.

In the multidimensional case the kinematic restriction also significantly affect the shape of the diffusion package (see. Fig.~\ref{fig:pdf_a01_d3}). In the considered case (three-dimensional walk with a point instantaneous source) at small $\alpha$ the asymptotic distribution of the particles is almost uniform. Indeed for small $\alpha$ the probability of the particles with long path is rather high. Therefore, a significant part of the particles is concentrated at the surface of the sphere $\mathbf{R}=\mathbf{v}t$. Since we study the distribution of $\mathbf{R}$ vector projection on the $x$ axis this leads to the formation of uniform distribution. With the increase of $\alpha$ the probability of a particle with long paths is reduced and so the role of scattering processes is increased. As a result, the particles leave the sphere $\mathbf{R}=\mathbf{v}t$ surface to the sphere interior leading to the formation of a "hump" in the vicinity of zero.

In the case of a finite mathematical expectation ($1<\alpha<2$) Eq.~(\ref{eq:m2a2}) shows that the diffusion package expands as $\Delta(t)\propto t^{\gamma(\alpha)}$ and $\gamma(\alpha)={(3-\alpha)/2}$. As you can see the package expands slower than $x=\pm vt$. This mitigates the effect of the kinematic restriction on the diffusion package formation, and the particle propagation corresponds to superdiffussion regime $\gamma(\alpha)\in(1/2,1)$. As shown in Ref.~\cite{Zolotarev1999}, in the case of walks with finite velocity and in the presence of traps the asymptotic distribution is described by a Levy stable law with the exponent factor $\alpha$. Accounting the final velocity is reduced to decrease of the diffusion coefficient $D\to D_v$.

\section{Conclusion}\label{sec:conclusion}

In the case of finiteness of the speed a motion analysis of the moments reveal appearance of three modes of the propagation during of the evolution of the process: ballistic, diffusive and asymptotic modes. The time of the transition between the ballistic and diffusive modes does not depend on $\alpha$ and dimension of the space. The time of transition from the diffusive mode to the asymptotic mode depends on $\alpha$. In the case of  $0<\alpha<1$ the time of reaching of the asymptotic mode rises with increasing of $\alpha$. At the fixed  $\alpha$ the moments of the different orders converge to their asymptotics at the equal time moments. Since kinematic restriction affect the formation of the diffusion packet as a result the asymptotics of the moments rises according to the ballistic law. In case ($1<\alpha<2$) the situation changes. In this case the main asymptotic term  is characterized by the dependence $M_{2n}^N(t)\propto (vt)^{2n+1-\alpha}$ and the kinematic restriction does not affect the formation of the particle distribution. The time of reaching of the asymptotic mode depends on $\alpha$ and the moments of the different orders reach their asymptotics at the different times. At the same time, the higher the moment order, the  earlier it reaches the  asymptotics. As a result the second moment has the most slow convergence to it asymptotic and this fact allows us to define the time of reaching the asymptotic mode as the time of the reaching of the second moment of it asymptotic. Accounting of the additional preasymptotic terms in the expressions for the $M_{2n}^N(t)$ decreases reaching time  the asymptotics in the comparison with the case when the main asymptotic term is only used.

The distribution of particles has been reconstructed by the method of moments. At $0<\alpha<1$  the kinematic restriction plays an important role both in the one-dimensional and three-dimensional cases. In one-dimensional case the asymptotic distributions are of $U$- and $W$-shapes. A similar result was obtained in \cite{Zolotarev1999, JarovikovaPhD2001_en, Uchaikin2003a, Zaburdaev2008, Uchaikin:JETPL:2010, Uchaikin2011, Uchaikin2011a, Uchaikin2014, Froemberg2015}. In addition, the fractional Laplacian in the fractional diffusion equation for finite velocity should be replaced by the material derivative of the fractional order \cite{Zaburdaev2002,Sokolov2003,Uchaikin2011,Uchaikin2011a}. In one-dimensional case, an analytical solution of this equation \cite{Uchaikin2011,Uchaikin2011a,Uchaikin2009,Froemberg2013a} is expressed through Lamperti distribution, however, for higher space dimensions the solution of this equation cannot be obtained. The method of moments provides a solution of this equation for any dimension. In particular, we have recovered the particle distribution in $x$-coordinates in the case of three-dimensional random walk. However, an accuracy of the recovery depends on the number of the moments used for the recovery.

At $1<\alpha<2$ recovering of the density through orthogonal polynomials was not successful. This is due to insufficient number of moments used for restoration, and as well due to their non-uniform convergence to the asymptotics. However, in this case, the asymptotic distribution of the particles is expressed through the density of the $N$-dimensional Levy stable law (see~(\ref{eq:solSupDfsDv})), and a finite velocity of the propagation could be accounted through the replacement of the diffusion coefficient $D\to D_v$.

\section{Acknowledgments}

The author expresses deep thanks to Borisova Christina for the help in the preparation of the English variant of the article manuscript.

\appendix
\section{The calculation of $\langle\cos^n\theta\rangle$}\label{app:cos}

Certain known formulas will be needed for the calculation of integral (\ref{eq:intcosn}). The full solid angle in the spherical coordinates in the multivariate space or which is just the same, the area of surface of a unit sphere in the multivariate space is defined as $ S_N=2\pi^{N/2}/\Gamma(N/2)$. The element of the solid angle in the multivariate space is $d\Omega_N=\prod_{k=1}^{N-1}\sin^{k-1}\theta_{k}d\theta_k$.
If we take it so that the angle $\theta$ is angle between the vector $\mathbf{R}_N(t-\xi/v)$ and the axis $x_N$ in the Cartesian coordinates then we obtain
\begin{align*}
\langle\cos^n\theta\rangle&=\frac{1}{S_N}\idotsint\limits_\Omega \cos^{n}\theta_{N-1}\sin\theta_2\sin^2\theta_3\dots\sin^{N-2}\theta_{N-1}d\theta_1 d\theta_2 d\theta_3\dots d\theta_{N-1}\\
&= \frac{S_{N-1}}{S_N}\int\limits_0^\pi\cos^n\theta_{N-1}\sin^{N-2}\theta_{N-1}d\theta_{N-1}.
\end{align*}
If we place under the sign of differential $\sin\theta_{N-1}$ and make the change of variable $\mu=\cos\theta_{N-1}$, we obtain
\begin{equation}\label{eq:intcosn1}
\langle\cos^n\theta\rangle=\frac{S_{N-1}}{S_N}\int\limits_{-1}^1\mu^n(1-\mu^2)^{(N-3)/2}d\mu.
\end{equation}
We are interested in the case of $n$ being an even number.

Further on, it will be necessary to apply to the following known integral (see integral 2.2.4.9 \citep{Prudnikov2002v1_eng})
$$
\int\limits_0^a x^{2m}(a^2-x^2)^{k-1/2}dx=\frac{a^{2m+2k}}{2}\frac{\Gamma(m+1/2)\Gamma(k+1/2)}{\Gamma(m+k+1)}.
$$
If we place expressions for $S_N$ and $S_{N-1}$ and this integral into Eq.~(\ref{eq:intcosn1}), we finally obtain
\begin{equation}\label{eq:meanCos}
\langle\cos^{2m}\theta\rangle=\frac{\Gamma(N/2)\Gamma(m+1/2)}{\sqrt{\pi}\Gamma(N/2+m)},
\end{equation}
where $m=n/2$.

\section{Expressions for the moments}\label{app:MomentsExpr}
\paragraph{The case $0<\alpha<1$}
\begin{align}
m_4^N(t)&=1/3\left(3-2\alpha+{\frac {\alpha}{N}}-\frac{\alpha^2}{N}\right)  \left(1-\alpha \right) (vt)^4,\label{eq:m4a01}\\
m_6^N(t)&=\left((5\alpha^2-17\alpha+15)N^2+3\alpha(2\alpha^2-5\alpha+3)N+2\alpha^2(\alpha^2-2\alpha+1)\right)
\frac{\left(1-\alpha \right)}{15N^2} (vt)^6,\\
m_8^N(t)&=\left((105+79\alpha^2-155\alpha-14\alpha^3)N^4+(210-28\alpha^4-12\alpha^2+92\alpha^3-232\alpha)N^3\right.-2\alpha(\alpha-1)\nonumber\\
\times&(10\alpha^3+7\alpha^2-89\alpha+87)N^2-\alpha^2 (5\alpha^2+43\alpha-72)(\alpha-1)^2N\left.-12\alpha^3(\alpha-1)^3\right)\frac{(\alpha-1)(vt)^8}{105N^3(N+2)}\\
%\end{align}
%\begin{align}
m_{10}^N(t)&=\left(48(\alpha-1)^4\alpha^4+2(7\alpha^2+113\alpha-180) (\alpha-1)^3N\alpha^3+ 5(14\alpha^3+53\alpha^2-259\alpha+228) (\alpha-1)^2N^2\alpha^2\right.\nonumber\\
&+5(\alpha-1)(27\alpha^4-46\alpha^3-195\alpha^2+562\alpha-390 )N^3\alpha+(42\alpha^4-344\alpha^3+1106\alpha^2-1644\alpha+95)N^5\nonumber\\
&\left.+(120\alpha^5-676\alpha^4+1132\alpha^3+282\alpha^2-2538\alpha+1890)N^4\right)\frac{(1-\alpha) (vt)^{10}}{945N^4(N+2)}.\label{eq:m10a01}
\end{align}

\paragraph{The case $1<\alpha<2$}
\begin{align}
M_4^N(t)&=\frac {4x_0^{\alpha-1} (\alpha-1)(vt)^{5-\alpha}}{\alpha(5-\alpha)(4-\alpha)}+
\frac {2(4+2N)(\alpha-1)^2x_0^{2\alpha-2}\left(\Gamma(2-\alpha)\right)^2(vt)^{6-2\alpha}}{N\alpha\Gamma(7-2\alpha)}\label{eq:m4a2}\\
M_6^N(t)&=\frac {6x_0^{\alpha-1}(\alpha-1)(vt)^{7-\alpha}}{\alpha(7-\alpha)(6-\alpha)}+
\frac{6(3N+12)x_0^{2\alpha-2}\Gamma(4-\alpha)\Gamma(2-\alpha)(1-\alpha)^2(vt)^{8-2\alpha}}
{N\alpha\Gamma(9-2\alpha)}\nonumber\\
&+\frac {2(3N+12)(4+2N)x_0^{3\alpha-3}\Gamma(2-\alpha)(\alpha-1)^3(vt)^{9-3\alpha}}{N^2\alpha\Gamma(10-3\alpha)}\\
%\end{align}
%\begin{align}
M_8^N(t)&=\frac {8x_0^{\alpha-1} (\alpha-1)(vt)^{9-\alpha}}{\alpha(9-\alpha)(8-\alpha)}+\left(4\Gamma(6-\alpha)
\Gamma(2-\alpha)+3{\frac {(N+4)\left(\Gamma(4-\alpha)\right)^2}{2+N}}\right)\frac{8(N+6)x_0^{2\alpha-2}(1-\alpha)^2(vt)^{10-2\alpha}}{N\alpha\Gamma(11-2\alpha)}\nonumber\\
&+96\frac{(N+6)(N+4)}{N^2}\frac{x_0^{3\alpha-3}\Gamma(4-\alpha) \left(\Gamma(2-\alpha)\right)^2(\alpha-1)^3(vt)^{11-3\alpha}} {\alpha\Gamma(12-3\alpha)}\nonumber\\
&+48\frac{(N+6)(N+4)(N+2)}{N^3}\frac{x_0^{4\alpha-4}\left(\Gamma(2-\alpha)\right)^4(1-\alpha)^4(vt)^{12-4\alpha}}{\alpha
\Gamma(13-4\alpha)}\\
%\end{align}
%\begin{align}
M_{10}^N(t)&=\frac {10x_0^{\alpha-1}(\alpha-1)(vt)^{11-\alpha}}{(11-\alpha)(10-\alpha)\alpha}+
\left(\frac{50(N+8)\Gamma(8-\alpha)\Gamma(2-\alpha)}{N}+\frac{100(N(14+N)+48)\Gamma(6-\alpha)\Gamma(4-\alpha)}{N(2+N)}\right)\nonumber\\
&\times\frac{x_0^{2\alpha-2}(1-\alpha)^2 (vt)^{12-2\alpha}}{\alpha\Gamma(13-2\alpha)}
+\left(\frac{40(N+8)(N+6)\left(4\Gamma(6-\alpha)(2+N)\Gamma(2-\alpha)
+3(N+4)\left(\Gamma(4-\alpha)\right)^2\right)}{N^2(2+N)}\right.\nonumber\\
&\left.+\frac{2(10N(14+N)+480)\left(2\Gamma(6-\alpha)(2+N)\Gamma(2-\alpha)+9(N+4)\left(\Gamma(4-\alpha)\right)^2\right)}
{N^2(2+N)}\right)\nonumber\\
&\times\frac{x_0^{3\alpha-3}(\alpha-1)^3\Gamma(2-\alpha)(vt)^{13-3\alpha}}{\alpha\Gamma(14-3\alpha)}
+\left( 480(N+8)(N+6)(N+4)+120(N(14+N)+48)(N+4)\right)\nonumber\\
&\times\frac{(1-\alpha)^4\left(\Gamma(2-\alpha)\right)^3\Gamma(4-\alpha) x_0^{4\alpha-4} (vt)^{14-4\alpha}}{\alpha\Gamma(15-4\alpha) N^3}\nonumber\\
&+240\frac{(N+8)(N+6)(N+4)(2+N)}{N^4}\frac{x_0^{5\alpha-5}(\alpha-1)^5\left(\Gamma(2-\alpha)\right)^5(vt)^{15-5\alpha}} {\alpha\Gamma(16-5\alpha)}\label{eq:m10a2}
\end{align}

\bibliographystyle{elsarticle-num}
\bibliography{d:/Bibliography/library}


%\begin{thebibliography}{00}


%\bibitem{}

%\end{thebibliography}
\end{document}